\documentclass[submission,copyright,creativecommons,sharealike,noncommercial]{eptcs}
\providecommand{\event}{QPL 2017}
\pdfoutput=1

\usepackage{mathtools} 
\usepackage{amssymb} 
\usepackage{bbold} 
\usepackage{amsthm} 
\usepackage{stmaryrd} 

\usepackage{relsize} 
\usepackage{microtype} 
\usepackage{csquotes} 

\usepackage{hyperref} 
\usepackage[nocompress]{cite} 

\usepackage{graphicx} 
\usepackage[usenames,dvipsnames]{xcolor} 
\usepackage{tikz} 
\usepackage{circuitikz} 
\usetikzlibrary{
	arrows,
	shapes,
	decorations,
	intersections,
	backgrounds,
	positioning,
	circuits.ee.IEC
	}

		\newcounter{theorem_c} 
		\numberwithin{equation}{section} 

		\theoremstyle{plain} 
		\newtheorem{theorem}[theorem_c]{Theorem}
		
		\newtheorem{lemma}[theorem_c]{Lemma}

		\newtheoremstyle{exampstyle}
		  {2mm} 
		  {2mm} 
		  {\itshape} 
		  {} 
		  {\bfseries} 
		  {.} 
		  {.5em} 
		  {} 

		\theoremstyle{exampstyle}
		\newtheorem{definition}[theorem_c]{Definition}

	\newcommand{\inlineQuote}[1]{\textquotedblleft #1\textquotedblright} 
	\newcommand{\goodvdots}{\protect\raisebox{7pt}{\vdots}} 

	\newcommand{\naturals}{\mathbb{N}} 
	\newcommand{\reals}{\mathbb{R}} 
	\newcommand{\complexs}{\mathbb{C}} 
	\newcommand{\restrict}[2]{\left. #1 \right\vert_{#2}} 
	
	\newcommand{\singletonSet}{\mathbb{1}}

	\newcommand{\suchthat}[2]{\left\{#1 \: \middle\vert \: #2\right\}} 

		\newcommand{\ket}[1]{\vert #1 \rangle} 
		\newcommand{\bra}[1]{\langle #1 \vert} 
		
		\newcommand{\Trace}[1]{\operatorname{Tr}\,#1} 
		\newcommand{\decohSym}{\operatorname{dec}} 
		\newcommand{\decoh}[1]{\decohSym_{#1}} 

		\newcommand{\SpaceH}{\mathcal{H}} 
		\newcommand{\SpaceG}{\mathcal{G}}
		\newcommand{\SpaceK}{\mathcal{K}}

		\newcommand{\id}[1]{id_{#1}} 

		\newcommand{\CMonCategory}{\operatorname{CMon}} 
		\newcommand{\RMatCategory}[1]{#1\operatorname{-Mat}} 
		\newcommand{\StochCategory}{\operatorname{Stoch}} 

		\newcommand{\fdHilbCategory}{\operatorname{fdHilb}} 
		\newcommand{\fRelCategory}{\operatorname{fRel}} 
		\newcommand{\fPFunCategory}{\operatorname{fPFun}} 
		\newcommand{\fSetCategory}{\operatorname{fSet}} 

		\newcommand{\CategoryC}{\mathcal{C}}
		\newcommand{\CategoryD}{\mathcal{D}}

		\newcommand{\obj}[1]{\operatorname{obj} \, #1} 
		\newcommand{\CPMCategory}[1]{\operatorname{CPM}[#1]} 
		\newcommand{\CPStarCategory}[1]{\operatorname{CP}^\ast[#1]} 

		\newcommand{\normalisedKaroubiEnvelope}[1]{\operatorname{Split}_{\!\hbox{\begin{tikzpicture} [scale=1.2,transform shape,rotate=-90] 

\def\deltax{0.3} 
\def\deltay{0.5} 

\path[use as bounding box] (-\deltax,-0.7*\deltay) rectangle (\deltax,0.3*\deltay);

\node (mult) at (0,0.3*\deltay) [upground,scale=0.5] {};
\node (mult_label_in) at (0,-0.7*\deltay) {};
\draw[-] (mult_label_in) to (mult);

\end{tikzpicture}
}\!}\left[#1\right]} 
		\newcommand{\decoheredSystemsSubcategory}[1]{\operatorname{Decoh}\left[#1\right]} 

		\newcommand{\classicalSubcategory}[1]{#1_{K}} 
		\newcommand{\classicalSystem}[1]{K_{#1}} 

	\newcommand{\classicalStates}[1]{K(#1)} 

	\newcommand{\Xbwcolour}{black!80}

	\newcommand{\Zbwcolour}{white}
	\newcommand{\ZbwdotSym}{\hbox{\begin{tikzpicture} [scale=1.2,transform shape] 

\def\deltax{0.3} 
\def\deltay{0.5} 

\node [dot, fill=\Zbwcolour] (mult) at (0,0) {};

\end{tikzpicture}
}\!\!} 

	\newcommand{\Ybwcolour}{black!15}

	\newcommand{\Wbwcolour}{black!55}

	\newcommand{\trace}[1]{\hbox{\begin{tikzpicture} [scale=1.2,transform shape,rotate=-90] 

\def\deltax{0.3} 
\def\deltay{0.5} 

\path[use as bounding box] (-\deltax,-0.7*\deltay) rectangle (\deltax,0.3*\deltay);

\node (mult) at (0,0.3*\deltay) [upground,scale=0.5] {};
\node (mult_label_in) at (0,-0.7*\deltay) {};
\draw[-] (mult_label_in) to (mult);

\end{tikzpicture}
}\!_{#1}} 

	\tikzset{
	  rectangle with rounded corners north west/.initial=4pt,
	  rectangle with rounded corners south west/.initial=4pt,
	  rectangle with rounded corners north east/.initial=4pt,
	  rectangle with rounded corners south east/.initial=4pt,
	}
	\makeatletter
	\pgfdeclareshape{rectangle with rounded corners}{
	  \inheritsavedanchors[from=rectangle] 
	  \inheritanchorborder[from=rectangle]
	  \inheritanchor[from=rectangle]{center}
	  \inheritanchor[from=rectangle]{north}
	  \inheritanchor[from=rectangle]{south}
	  \inheritanchor[from=rectangle]{west}
	  \inheritanchor[from=rectangle]{east}
	  \inheritanchor[from=rectangle]{north east}
	  \inheritanchor[from=rectangle]{south east}
	  \inheritanchor[from=rectangle]{north west}
	  \inheritanchor[from=rectangle]{south west}
	  \backgroundpath{
	    \southwest \pgf@xa=\pgf@x \pgf@ya=\pgf@y
	    \northeast \pgf@xb=\pgf@x \pgf@yb=\pgf@y
	    \pgfkeysgetvalue{/tikz/rectangle with rounded corners north west}{\pgf@rectc}
	    \pgfsetcornersarced{\pgfpoint{\pgf@rectc}{\pgf@rectc}}
	    \pgfpathmoveto{\pgfpoint{\pgf@xa}{\pgf@ya}}
	    \pgfpathlineto{\pgfpoint{\pgf@xa}{\pgf@yb}}
	    \pgfkeysgetvalue{/tikz/rectangle with rounded corners north east}{\pgf@rectc}
	    \pgfsetcornersarced{\pgfpoint{\pgf@rectc}{\pgf@rectc}}
	    \pgfpathlineto{\pgfpoint{\pgf@xb}{\pgf@yb}}
	    \pgfkeysgetvalue{/tikz/rectangle with rounded corners south east}{\pgf@rectc}
	    \pgfsetcornersarced{\pgfpoint{\pgf@rectc}{\pgf@rectc}}
	    \pgfpathlineto{\pgfpoint{\pgf@xb}{\pgf@ya}}
	    \pgfkeysgetvalue{/tikz/rectangle with rounded corners south west}{\pgf@rectc}
	    \pgfsetcornersarced{\pgfpoint{\pgf@rectc}{\pgf@rectc}}
	    \pgfpathclose
	 }
	}
	\makeatother

	\tikzset{->-/.style={decoration={markings,mark=at position #1 with {\arrow{>}}},postaction={decorate}}}
	\tikzset{-<-/.style={decoration={markings,mark=at position #1 with {\arrow{<}}},postaction={decorate}}}

	\tikzstyle{every picture}=[baseline=-0.25em,scale=0.5]
	\pgfdeclarelayer{edgelayer}
	\pgfdeclarelayer{nodelayer}
	\pgfsetlayers{edgelayer,nodelayer,main}

	\tikzstyle{box} = [draw,shape=rectangle,inner sep=2pt,minimum height=6mm,minimum width=6mm,fill=white] 
	\tikzstyle{boxlarge} = [draw,shape=rectangle,inner sep=2pt,minimum height=1.5cm,minimum width=8mm,fill=white] 
	\tikzstyle{boxLarge} = [draw,shape=rectangle,inner sep=2pt,minimum height=2cm,minimum width=10mm,fill=white] 
	\tikzstyle{boxsmall} = [draw,shape=rectangle,inner sep=2pt,minimum height=3mm,minimum width=3mm,fill=white] 
	\tikzstyle{dot} = [inner sep=0mm,minimum width=3mm,minimum height=3mm,draw,shape=circle,text depth=-0.1mm]
	\tikzstyle{Zbwdot} = [dot, fill=\Zbwcolour]
	\tikzstyle{Xbwdot} = [dot, fill=\Xbwcolour]
	\tikzstyle{Ybwdot} = [dot, fill=\Ybwcolour]
	\tikzstyle{Wbwdot} = [dot, fill=\Wbwcolour]
	\tikzstyle{antipode} = [boxsmall] 

	\tikzstyle{state} = [draw, rectangle with rounded corners,
	  rectangle with rounded corners north west=8pt,
	  rectangle with rounded corners south west=8pt,
	  rectangle with rounded corners north east=0pt,
	  rectangle with rounded corners south east=0pt,
	,inner sep=2pt,minimum height=6mm,minimum width=6mm,fill=white]
	\tikzstyle{statelarge} = [draw, rectangle with rounded corners,
	  rectangle with rounded corners north west=8pt,
	  rectangle with rounded corners south west=8pt,
	  rectangle with rounded corners north east=0pt,
	  rectangle with rounded corners south east=0pt,
	,inner sep=2pt,minimum height=1.5cm,minimum width=8mm,fill=white]
	\tikzstyle{stateLarge} = [draw, rectangle with rounded corners,
	  rectangle with rounded corners north west=8pt,
	  rectangle with rounded corners south west=8pt,
	  rectangle with rounded corners north east=0pt,
	  rectangle with rounded corners south east=0pt,
	,inner sep=2pt,minimum height=2cm,minimum width=8mm,fill=white]
	\tikzstyle{effect} = [draw, rectangle with rounded corners,
	  rectangle with rounded corners north west=0pt,
	  rectangle with rounded corners south west=0pt,
	  rectangle with rounded corners north east=8pt,
	  rectangle with rounded corners south east=8pt,
	,inner sep=2pt,minimum height=6mm,minimum width=6mm,fill=white]
	\tikzstyle{scalar}=[diamond,draw,inner sep=1pt,font=\small,fill=white]

	\tikzstyle{cdnode}=[fill=white]
	\tikzstyle{labelnode}=[fill=white]
	\tikzstyle{tightlabelnode}=[fill=white,inner sep = 0.1mm]
	\tikzstyle{none}=[inner sep=0pt]
	\tikzstyle{whiteline}=[-, line width=4pt, draw=white]

	\tikzstyle{trace}=[circuit ee IEC,thick,ground,scale=2.5]
	\tikzstyle{cotrace}=[circuit ee IEC,thick,ground,rotate=180,scale=2.5]
	\tikzstyle{upground}=[circuit ee IEC,thick,ground,rotate=90,scale=2.5]
	\tikzstyle{downground}=[circuit ee IEC,thick,ground,rotate=-90,scale=2.5]

	\tikzstyle{doubled} = [line width=1.8pt] 
	
	\tikzstyle{empty diagram}=[draw=gray!40!white,dashed,shape=rectangle,minimum width=1cm,minimum height=1cm]

\title{Categorical Probabilistic Theories}
\author{
	Stefano Gogioso\\
	University of Oxford \\
	\texttt{stefano.gogioso@cs.ox.ac.uk}
	\and
	Carlo Maria Scandolo \\
	University of Oxford \\
	\texttt{carlomaria.scandolo@cs.ox.ac.uk}
}

\def\titlerunning{Categorical Probabilistic Theories}
\def\authorrunning{S. Gogioso \& C. M. Scandolo}

\begin{document}

\maketitle

\begin{abstract}
	We present a simple categorical framework for the treatment of probabilistic theories, with the aim of reconciling the fields of Categorical Quantum Mechanics (CQM) and Operational Probabilistic Theories (OPTs). In recent years, both CQM and OPTs have found successful application to a number of areas in quantum foundations and information theory: they present many similarities, both in spirit and in formalism, but they remain separated by a number of subtle yet important differences. We attempt to bridge this gap, by adopting a minimal number of operationally motivated axioms which provide clean categorical foundations, in the style of CQM, for the treatment of the problems that OPTs are concerned with. 
\end{abstract}

\section{Introduction} 
\label{section_introduction}

Categorical methods are finding more and more applications in the foundations of quantum theory and quantum information, with two main frameworks currently dominating the scene: Categorical Quantum Mechanics (CQM) and Operational Probabilistic Theories (OPTs). The CQM framework \cite{Abramsky2009,Coecke2011,Backens2014,Coecke2015,Coecke2016a} concerns itself with the compositional and algebraic structure of quantum system and processes, and has a heavy focus on diagrammatic methods \cite{Coecke2009p,Horsman2011,Kissinger2012}. The OPT framework \cite{Chiribella2010,Chiribella2011,hardy2011,Hardy-informational-2,Chiribella2014,hardy2013,Chiribella2015b} was developed with the aim of obtaining a characterisation of quantum theory in terms of information-theoretic axioms, inside the wider community of generalised probabilistic theories \cite{Hardy2001,Barrett2007,Barnum-1,Brukner,Masanes-physical-derivation,Barnum-2}. 

Both frameworks are based on symmetric monoidal categories, and at first sight it looks like they should be complementing each other. However, a number of subtle differences have resulted in a significant disconnect between the two communities, with duplication of efforts and loss of synergy. The OPT framework is founded on explicit probabilistic structure, axioms of concrete physical inspiration and traditional proof methods, which make it hard to work with in a categorical and diagrammatic setting. Conversely, the CQM framework has clean categorical foundations and diagrammatic proof methods, but at the expense of direct physical interpretation. 

We set off to reconcile the two frameworks, by proposing categorical foundations in the style of CQM for the kind of systems and results that are of interest for the OPT community. We define a probabilistic theory as a symmetric monoidal category satisfying three additional requirements: the existence of classical systems, the existence of probabilistic structure, and the possibility of defining local states. Our requirements are formulated in standard categorical terms, and can be readily adopted within the CQM framework; at the same time, they correspond to relatable physical and operational requirements that any probabilistic theory should possess. The end result is a framework which is simple and rigorous in its foundations, shows direct operational significance, and comes with more expressive power than either of the CQM or OPT frameworks alone. 

The categorical probabilistic theories defined in this work provide the underlying framework for an upcoming novel derivation of quantum theory from diagrammatic postulates \cite{Selby2017}. The framework is also expressive enough to model many toy theories of current interest in the foundations of quantum theory \cite{Gogioso2017d}, such as real, hyperbolic, relational and modal quantum theory.

Finally, we should mention that other efforts to connect the CQM and OPT communities have appeared recently in the literature: some of the most notable ones include Ref. \cite{Tull2016}, coming from the logical perspective of effectus theory, Ref. \cite{Tull2017}, in relation to leaks, decoherence and the emergence of classicality, and Ref. \cite{Fritz2014}, in relation to causality and Bell-type measurement scenarios.

\section{Classical Theory} 
\label{section_CT}

The main ingredient distinguishing OPTs from the categories used in the context of CQM is the explicit presence of probabilistic mixtures, or equivalently the existence of classical systems taking part in the processes. From the point of view of OPTs, these classical systems are absorbed into the formalism, using indices and summations. From the point of view of CQM, on the other hand, the only distinction between classical and quantum systems is operational, and both should be modelled explicitly by a theory. 

\subsection{Classical theory as a SMC}

Classical systems\footnote{This work, like most works in OPTs and CQM, is concerned with \textit{finite} classical systems alone.} and stochastic processes between them form a symmetric monoidal category (SMC), usually known as $\StochCategory$. The objects of $\StochCategory$ are labelled by finite sets, and the morphisms $X \rightarrow Y$ in $\StochCategory$ are the $Y$-by-$X$ stochastic matrices, with matrix composition as sequential composition and the Kronecker product as parallel composition (aka tensor product). 

From a categorical standpoint, the restriction to convex combinations required by $\StochCategory$ is inconvenient: if we wish to use a summation operation, as is usually done in the treatment of classical systems and mixed-state quantum theory, we have to introduce external restrictions on the kind of sums that we allow. Alternatively, we could see the summation as defined on a larger category, with $\StochCategory$ arising as the sub-category of suitably \textit{normalised} processes. This latter approach is in line with the general philosophy of CQM: one studies a broader collection of abstract processes, easier to describe and to manipulate, and then assembles/restricts them appropriately to obtain the subcategory of concrete, physically relevant processes. As our model of classical systems and (potentially non-normalised) non-deterministic processes between them, we take the SMC $\RMatCategory{\reals^+}$: just like $\StochCategory$, the objects are the finite non-empty sets, but the morphisms $X \rightarrow Y$ are ALL the $\reals^+$-valued matrices in $\big(\reals^+\big)^{Y \times X}$, not just the stochastic ones. We refer to these maps as the \textbf{probabilistic processes}, and use \textbf{normalised} to denote the stochastic ones.

\subsection{Linear structure}

We can equip $\RMatCategory{\reals^+}$ with a well-defined notion of summation by suitably enriching it in commutative monoids: we endow the morphisms $X \rightarrow Y$ with the linear structure of $\big(\reals^+\big)^{Y \times X}$, i.e. we take matrix addition as our summation operation and the zero matrix as its neutral element (modelling the impossible process). Traditionally, the only requirement for a category enriched in commutative monoids is that sequential composition be linear\footnote{Specifically, we require $f \circ (g + h) = (f \circ g) + (f \circ h)$, $ (g + h) \circ f = (g \circ f) + (h \circ f)$, $0 \circ f = 0$ and $f \circ 0 = 0$.}, but in SMCs it makes sense to further require that parallel composition be linear as well\footnote{Things are a bit more complicated with the tensor structure: the tensor product is linear, just like composition, but we also need to require the associator and unitors to be linear.}. This is indeed the case for our choice of enrichment.

\subsection{Deterministic processes}

A special place amongst the many probabilistic processes between two classical systems $X$ and $Y$ is held by the \textbf{deterministic processes}\footnote{In line with its common meaning in computer science, we use the word \textbf{deterministic} to denote classical processes which map each definite input to a single definite output. We use \textbf{stochastic} (or normalised probabilistic, in this work) to denote classical processes which more generally map definite inputs to probability distributions over definite outputs.}, the functions $X \rightarrow Y$. The deterministic processes form the sub-SMC $\fSetCategory$ of $\RMatCategory{\reals^+}$, and the Kronecker product restricts to the Cartesian product on them. The states in $\fSetCategory$ for a classical system $X$ are its \textbf{deterministic states}, the probability distributions $\delta_{x}$ concentrated entirely on a single point $x \in X$ (or, if you prefer, the linear extensions of the points of $X$ themselves). The singleton set $\singletonSet := \{1\}$ is the terminal object in $\fSetCategory$, and we refer to the unique effect $X \rightarrow \singletonSet$ as the \textbf{discarding map}\footnote{In a not-at-all-unexpected coincidence, it makes total sense to call this \textbf{the deterministic effect}.} on the classical system $X$, which is graphically denoted by a ground symbol $\trace{X}$. The deterministic processes sit inside a larger sub-SMC $\fPFunCategory$ of \textbf{partial deterministic processes}, with partial functions as morphisms. 

\subsection{Resolution of the identity}

In order to clearly distinguish classical systems from more general systems in the theory, we will adopt a special graphical convention. We use dashed wires to denote systems which are guaranteed to be classical, so that the identity morphism $X \rightarrow X$ for a classical system $X$ will be represented graphically as follows:
\begin{equation}\label{idClassical}
\begin{tikzpicture}
	\begin{pgfonlayer}{nodelayer}
		\node [style=labelnode] (0) at (-2, 0) {$X$};
		\node [style=labelnode] (1) at (2, 0) {$X$};
	\end{pgfonlayer}
	\begin{pgfonlayer}{edgelayer}
		\draw [style=dashed] (0) to (1);
	\end{pgfonlayer}
\end{tikzpicture}
\end{equation}
Using the enrichment, we obtain the following resolution of the identity: 
\begin{equation}\label{idClassicalResolution}
\begin{tikzpicture}
	\begin{pgfonlayer}{nodelayer}
		\node [style=labelnode] (0) at (-6, 0) {$X$};
		\node [style=labelnode] (1) at (-2, 0) {$X$};
		\node [style=labelnode] (2) at (4, 0) {$X$};
		\node [style=labelnode] (3) at (10, 0) {$X$};
		\node [style=labelnode] (4) at (0, 0) {$=$};
		\node [style=labelnode] (5) at (2, -0.25) {$\sum\limits_{x \in X}$};
		\node [style=effect] (6) at (6.25, 0) {$x$};
		\node [style=state] (7) at (7.75, 0) {$x$};
	\end{pgfonlayer}
	\begin{pgfonlayer}{edgelayer}
		\draw [style=dashed] (0) to (1);
		\draw [style=dashed] (2) to (6);
		\draw [style=dashed] (7) to (3);
	\end{pgfonlayer}
\end{tikzpicture}
\end{equation}
The state labelled by $x \in X$ above is the deterministic state corresponding to $x$; the effect labelled by $x$ above is the one sending $x$ to $1\in \reals^+$ and all $x' \neq x$ to $0 \in \reals^+$ (or, if you prefer, the linear extension of the partial function $X \rightharpoonup \singletonSet$ which sends $x$ to $1 \in \singletonSet$ and is undefined on all other $x' \in X$). 

\subsection{Marginalisation}

Because the resolution of the identity is in terms of deterministic states, we immediately obtain the following resolution of the discarding map:
\begin{equation}\label{discardingMapClassicalResolution}
\begin{tikzpicture}
	\begin{pgfonlayer}{nodelayer}
		\node [style=labelnode] (0) at (-5, 0) {$X$};
		\node [style=trace] (1) at (-2, 0) {};
		\node [style=labelnode] (2) at (4, 0) {$X$};
		\node [style=labelnode] (3) at (0, 0) {$=$};
		\node [style=labelnode] (4) at (2, -0.25) {$\sum\limits_{x \in X}$};
		\node [style=effect] (5) at (6.25, 0) {$x$};
	\end{pgfonlayer}
	\begin{pgfonlayer}{edgelayer}
		\draw [style=dashed] (0) to (1);
		\draw [style=dashed] (2) to (5);
	\end{pgfonlayer}
\end{tikzpicture}
\end{equation}
This means that discarding a subsystem in $\RMatCategory{\reals^+}$ corresponds to the usual notion of \textbf{marginalisation}. From the CQM perspective, the process of localisation of states and processes in arbitrary SMCs is captured by the notion of \textbf{environment structure} \cite{Coecke2013a,Coecke2016}. Because classical states and processes are localised by marginalising, it is no surprise that discarding maps in $\RMatCategory{\reals^+}$ satisfy the requirements to form an environment structure:
\begin{equation}\label{classicalEnvironmentStructure}
\begin{tikzpicture}[scale=0.8]
	\begin{pgfonlayer}{nodelayer}
		\node [style=labelnode] (0) at (-7.5, 0) {$=$};
		\node [style=labelnode, inner sep=0.1 mm] (1) at (-13.5, 0) {$X\otimes Y$};
		\node [style=trace] (2) at (-9.5, 0) {};
		\node [style=labelnode, inner sep=0.1 mm] (3) at (-5.5, 0.75) {$X$};
		\node [style=trace] (4) at (-2.5, 0.75) {};
		\node [style=labelnode, inner sep=0.1 mm] (5) at (-5.5, -0.75) {$Y$};
		\node [style=trace] (6) at (-2.5, -0.75) {};
		\node [style=labelnode, inner sep=0.1 mm] (7) at (2.5, 0) {$\singletonSet$};
		\node [style=trace] (8) at (5.5, 0) {};
		\node [style=labelnode] (9) at (7.5, 0) {$=$};
		\node [style=empty diagram] (9) at (10.5, 0) {};
	\end{pgfonlayer}
	\begin{pgfonlayer}{edgelayer}
		\draw [style=dashed] (1) to (2);
		\draw [style=dashed] (3) to (4);
		\draw [style=dashed] (5) to (6);
		\draw [style=dashed] (7) to (8);
	\end{pgfonlayer}
\end{tikzpicture}
\end{equation}
The empty diagram on the right hand side of the second equation is the scalar $1$, which as a diagrammatic convention is usually omitted. 

\subsection{Normalised processes}

A choice of environment structure always singles out a sub-SMC of \textbf{normalised} processes, namely those processes $f$ satisfying the following condition:
\begin{equation}\label{normalisedClassical}
\begin{tikzpicture}[scale=0.8]
	\begin{pgfonlayer}{nodelayer}
		\node [style=box] (0) at (-5.5, 0) {$f$};
		\node [style=labelnode, inner sep=0.1 mm] (1) at (-8.5, 0) {};
		\node [style=trace] (2) at (-2, 0) {};
		\node [style=labelnode] (3) at (0, 0) {$=$};
		\node [style=labelnode, inner sep=0.1 mm] (4) at (2, 0) {};
		\node [style=trace] (5) at (6, 0) {};
	\end{pgfonlayer}
	\begin{pgfonlayer}{edgelayer}
		\draw [style=dashed] (1) to (0);
		\draw [style=dashed] (0) to (2);
		\draw [style=dashed] (4) to (5);
	\end{pgfonlayer}
\end{tikzpicture}
\end{equation}
The abstract normalisation condition above allows us to recover $\StochCategory$ in a categorically fashionable way, as the sub-SMC of normalised processes in $\RMatCategory{\reals^+}$. As a bonus, we also get $\fSetCategory$ as the sub-SMC of normalised processes in $\fPFunCategory$.

\subsection{Probabilities as a resource}

In computer science, it is common for commutative semirings to model resources used in computation: for example, the \textit{probability semiring} $(\reals^{+},+,\times)$ is used to model probabilistic computation, the \textit{boolean semiring} $(\mathbb{B},\vee,\wedge)$ is used to model existence problems, and the \textit{natural numbers} $(\naturals,+,\times)$ are used to model counting problems\footnote{One could also mention the \textit{tropical semiring} $(\reals, \min , + )$ used in the Floyd-Warshall algorithm, or the \textit{Viterbi semiring} $([0,1], \max, \times)$ used in the Viterbi algorithm, but also the $p$-adic numbers and finite fields.}. A similar approach could be adopted in physics: probabilities (corresponding to the semiring $\reals^+$) could be seen as a resource providing non-determinism in classical systems, and we might wish to study the toy theories obtained by using other resources (corresponding to other semirings) in their place. This is, for example, the road taken by the sheaf-theoretic framework of Ref. \cite{Abramsky2011}, where a number of important results on non-locality and contextuality are obtained by confronting the probabilities (the semiring $\reals^+$) with possibilities \cite{Logical-nonlocality} (the semiring $\mathbb{B}$) and signed probabilities \cite{Negative-probabilities-1,Negative-probabilities-2} (the semiring $\reals$, which is given an operational interpretation in Ref. \cite{Abramsky2014}). The computational complexity implications of different choices of semirings are explored in \cite{DeBeaudrap2014}.

If we use a generic commutative semiring $R$ to model non-determinism, the SMC of classical systems and \textbf{$R$-probabilistic} processes between them is given by the category $\RMatCategory{R}$ of $R$-valued matrices: we refer to the normalised states as \textbf{$R$-distributions}, and to the normalised processes as \textbf{$R$-stochastic}. As an example, $\RMatCategory{\mathbb{B}}$ is the category $\fRelCategory$ of finite sets and relations between them, a well-studied toy model in CQM \cite{Coecke2008,Coecke2012a,Gogioso2015b}. All the theory we have developed above for $\RMatCategory{\reals^+}$ straightforwardly extends to general semirings, as long as we replace $\StochCategory$ with the appropriate notion of normalised processes.  The categories $\fSetCategory$ and $\fPFunCategory$ of total and partial deterministic processes are always sub-SMCs of $\RMatCategory{R}$, showing that our interpretation of $R$ as modelling a notion of non-determinism is sound.

\section{Probabilistic Theories}
\label{section_CPTs}

When talking about a \textit{theory}, we broadly mean a categorical model (a SMC) that captures physical systems and the compositional structure of processes between them. In the spirit of CQM, we avoid restricting our attention to physical processes alone, but instead we allow the presence of a whole spectrum of idealised, abstract processes that provide building blocks for physical processes, or otherwise help in reasoning about them. Contrary to the OPT framework, we take the view that a theory should specify objects and processes in a purely compositional way, and need not necessarily make immediate reference to probabilistic outcomes of tests and measurement: as a consequence, we do not take any quotient with respect to probabilistic outcomes, and we allow for the possibility that a theory specifies processes which are different but cannot be distinguished by means of tests or measurements alone (this means that some alternative \inlineQuote{hidden variable theories} with the same operational predictions can be modelled).

\subsection{Probabilistic theories}

When talking about a \textit{probabilistic theory}, we mean a SMC which includes at least the classical systems amongst its ranks, and which is furthermore compatible with a couple of basic operational features, namely their probabilistic structure and marginalisation. We briefly motivate our requirements below.
\begin{enumerate}
	\item Every probabilistic theory has classical probabilistic systems under the hood: because these are themselves physical systems, we model them explicitly. In particular, their interface with other systems (e.g. measurements and preparations) can be talked about in compositional terms.
	\item It makes no sense to talk about a probabilistic theory if the probabilistic structure does not extend from classical systems to arbitrary systems. If this were not the case, one would not necessarily be able to work with scenarios in which multiplexed processes are controlled by a classical random variable, or to condition a process based on a classical output.
	\item Every probabilistic theory with operational aspirations should include a notion of localisation of states and processes. Indeed, the absence of a notion of local state compatible with marginalisation renders most protocol specifications meaningless, a fate they share with the notion of no-signalling and with the probabilistic foundations of thermodynamics.\footnote{In absence of a canonical notion of local state, talking about a localised computation or experimental setup requires non-trivial amounts of information about the state of the universe to be known. Sensible theories that do this exist, e.g. Everettian quantum theory, but they present a number of operational challenges and are not probabilistic in nature.}
\end{enumerate}
We now proceed to formalise these requirements in categorical terms.
\begin{definition}[\textbf{Probabilistic Theory}]\hfill\\
A \textbf{probabilistic theory} is a symmetric monoidal category $\CategoryC$ which satisfies the following requirements.
\begin{enumerate}
	\item There is a full sub-SMC of $\CategoryC$, which we denote by $\classicalSubcategory{\CategoryC}$, which is equivalent to $\RMatCategory{\reals^+}$. 
	\item The SMC $\CategoryC$ is enriched in commutative monoids\footnote{This means that for any two objects $\SpaceH$ and $\SpaceK$, the set of morphisms $\SpaceH \rightarrow \SpaceK$ comes with the structure of a commutative monoid, i.e. it comes with a commutative associative binary operation $+$ and a neutral element $0$ for it. Furthermore, composition of morphisms and tensor product of morphisms are both bilinear operations (e.g. $(f+g)\circ h = f\circ h + g \circ h$, $f \otimes 0 = 0$, etc).} and the enrichment on $\classicalSubcategory{\CategoryC}$ coincides with the one given by the linear structure of $\RMatCategory{\reals^+}$\footnote{Requirement 1 above imposes that $\classicalSubcategory{\CategoryC}$ be equivalent to $\RMatCategory{\reals^+}$, but says nothing about addition $+$ of morphisms and $0$ morphisms being respected by the equivalence. Requirement 2 further imposes that the equivalence between the two categories be a linear functor, so that the commutative monoid structure on morphisms is respected.}.
	\item The SMC $\CategoryC$ comes with an environment structure, i.e. with a family $(\trace{\SpaceH}:\SpaceH \rightarrow \singletonSet)_{\SpaceH \in \obj{\CategoryC}}$ of morphisms which satisfy the following requirements:
		\begin{equation}\label{environmentStructure}
		\begin{tikzpicture}[scale=0.8]
	\begin{pgfonlayer}{nodelayer}
		\node [style=labelnode] (0) at (-7.5, 0) {$=$};
		\node [style=labelnode, inner sep=0.1 mm] (1) at (-13.5, 0) {$\SpaceH \otimes \SpaceG$};
		\node [style=trace] (2) at (-9.5, 0) {};
		\node [style=labelnode, inner sep=0.1 mm] (3) at (-5.5, 0.75) {$\SpaceH$};
		\node [style=trace] (4) at (-2.5, 0.75) {};
		\node [style=labelnode, inner sep=0.1 mm] (5) at (-5.5, -0.75) {$\SpaceG$};
		\node [style=trace] (6) at (-2.5, -0.75) {};
		\node [style=labelnode, inner sep=0.1 mm] (7) at (2.5, 0) {$\singletonSet$};
		\node [style=trace] (8) at (5.5, 0) {};
		\node [style=labelnode] (9) at (7.5, 0) {$=$};
		\node [style=empty diagram] (9) at (10.5, 0) {};
	\end{pgfonlayer}
	\begin{pgfonlayer}{edgelayer}
		\draw [style=-] (1) to (2);
		\draw [style=-] (3) to (4);
		\draw [style=-] (5) to (6);
		\draw [style=-] (7) to (8);
	\end{pgfonlayer}
\end{tikzpicture}
		\end{equation}
	The environment structure on $\classicalSubcategory{\CategoryC}$ coincides with the one given by the discarding maps of $\RMatCategory{\reals^+}$.
\end{enumerate}
The requirements above can be generalised by replacing $\reals^+$ with an arbitrary commutative semiring $R$, in which case we will talk about an \textbf{$R$-probabilistic theory}.
\end{definition}
\noindent In the context of a specific $R$-probabilistic theory, we refer to $\classicalSubcategory{\CategoryC}$ as the \textbf{classical theory}, to its object as the \textbf{classical systems} and to its morphisms as the \textbf{classical processes}. In particular, $\singletonSet$ is the tensor unit for $\CategoryC$, and the scalars of $\CategoryC$ form the semiring $R$. For each pair of objects $\SpaceH, \SpaceK \in \obj{\CategoryC}$, the processes $\SpaceH \rightarrow \SpaceK$ in $\CategoryC$ form a commutative monoid, with a summation operation $+$ and a zero element $0$, which we refer to as the \textbf{impossible process}. In line with the nomenclature adopted for classical systems, we refer to the $\trace{\SpaceH}$ maps involved in the environment structure as the \textbf{discarding maps}, and to those processes $f$ satisfying the following equation as \textbf{normalised}:
\begin{equation}\label{normalised}
\begin{tikzpicture}[scale=0.8]
	\begin{pgfonlayer}{nodelayer}
		\node [style=box] (0) at (-5.5, 0) {$f$};
		\node [style=labelnode, inner sep=0.1 mm] (1) at (-8.5, 0) {};
		\node [style=trace] (2) at (-2, 0) {};
		\node [style=labelnode] (3) at (0, 0) {$=$};
		\node [style=labelnode, inner sep=0.1 mm] (4) at (2, 0) {};
		\node [style=trace] (5) at (6, 0) {};
	\end{pgfonlayer}
	\begin{pgfonlayer}{edgelayer}
		\draw [-] (1) to (0);
		\draw [-] (4) to (5);
		\draw [style=-] (0) to (2);
	\end{pgfonlayer}
\end{tikzpicture}
\end{equation}


\subsection{Tests}

Consider a process $f$ with a classical output valued in some finite set $Y$:
\begin{equation}\label{processWithClassOutput}
\begin{tikzpicture}
	\begin{pgfonlayer}{nodelayer}
		\node [style=box] (0) at (0, 0) {$f$};
		\node [style=tightlabelnode] (1) at (3, -0.75) {$Y$};
		\node [style=none] (2) at (0.5, -0.25) {};
		\node [style=tightlabelnode] (3) at (3, 0.25) {$\SpaceG$};
		\node [style=tightlabelnode] (4) at (-3, 0) {$\SpaceH$};
		\node [style=none] (5) at (0.5, 0.25) {};
	\end{pgfonlayer}
	\begin{pgfonlayer}{edgelayer}
		\draw [style=dashed, in=180, out=0, looseness=1.25] (2.center) to (1);
		\draw [style=-] (4) to (0);
		\draw [style=-] (3) to (5.center);
	\end{pgfonlayer}
\end{tikzpicture}
\end{equation}
Operationally, we can think of each output value $y \in Y$ as corresponding to a different process being performed\footnote{A value $y$ which is never output can, without loss of generality, be thought to correspond to the impossible process.}, and that process can be obtained by \textbf{testing against} the output value $y$: 
\begin{equation}\label{testingAgainstOutput}
\begin{tikzpicture}
	\begin{pgfonlayer}{nodelayer}
		\node [style=effect] (0) at (6.5, -0.75) {$y$};
		\node [style=none] (1) at (4.5, -0.25) {};
		\node [style=box] (2) at (4, 0) {$f$};
		\node [style=tightlabelnode] (3) at (1.5, 0) {$\SpaceH$};
		\node [style=none] (4) at (4.5, 0.25) {};
		\node [style=tightlabelnode] (5) at (8, 0.25) {$\SpaceG$};
		\node [style=tightlabelnode] (6) at (0, 0) {$:=$};
		\node [style=box] (7) at (-4, 0) {$f_y$};
		\node [style=tightlabelnode] (8) at (-6.5, 0) {$\SpaceH$};
		\node [style=tightlabelnode] (9) at (-1.5, 0) {$\SpaceG$};
	\end{pgfonlayer}
	\begin{pgfonlayer}{edgelayer}
		\draw [style=dashed, in=180, out=0, looseness=1.50] (1.center) to (0);
		\draw [style=-] (3) to (2);
		\draw [style=-] (5) to (4.center);
		\draw (8) to (9);
	\end{pgfonlayer}
\end{tikzpicture}
\end{equation}
It should be noted that, even when $f$ is normalised, the process $f_y$ defined in Equation \ref{testingAgainstOutput} is not necessarily so, instead being weighted by the probability of $y$ occurring. In this sense, a process in the form of \ref{processWithClassOutput} corresponds to the notion of a \textbf{test} in OPTs \cite{Chiribella2010,Chiribella2015b}; if the input system $\SpaceH$ is trivial then it is a \textbf{preparation test}, while if the output system $\SpaceK$ is trivial then it is an \textbf{observation test}. Note that in this framework the difference between a test and a more general process is entirely in the eye of the beholder: when saying \inlineQuote{test} instead of \inlineQuote{process}, we will mean that we are mainly interested in its classical outputs, but we will not exclude the presence of classical systems amongst the input systems of the process.

\subsection{Output probability}

The \textbf{$R$-probability}\footnote{We use the word $R$-probability to denote the concept corresponding to probabilities when the probability semiring $\reals^+$ is replaced by a generic commutative semiring $R$: as the latter is fixed by the choice of $R$-probabilistic theory, this should cause no confusion. In specific applications, one might want to use more specialised terms, such as: \textbf{probability} for $R = \reals^+$; \textbf{signed probability} for $R=\reals$; \textbf{possibility} for $R = \mathbb{B}$; \textbf{count} for $R = \naturals$.} $\mathbb{P}_{\!\rho}(y)$ of output value $y \in Y$ occurring in a preparation test can be obtained as follows:
\begin{equation}\label{outcomeProbabilityPrepTest}
\begin{tikzpicture}
	\begin{pgfonlayer}{nodelayer}
		\node [style=effect] (0) at (4.5, -0.75) {$y$};
		\node [style=none] (1) at (2.5, -0.25) {};
		\node [style=state] (2) at (2, 0) {$\rho$};
		\node [style=none] (3) at (2.5, 0.25) {};
		\node [style=trace] (4) at (6, 0.25) {};
		\node [style=tightlabelnode] (5) at (0, 0) {$:=$};
		\node [style=tightlabelnode] (6) at (-1.5, 0) {$\mathbb{P}_{\!\rho}(y)$};
	\end{pgfonlayer}
	\begin{pgfonlayer}{edgelayer}
		\draw [style=dashed, in=180, out=0, looseness=1.50] (1.center) to (0);
		\draw [style=-] (4) to (3.center);
	\end{pgfonlayer}
\end{tikzpicture}
\end{equation}
When the $R$-probability of outcome $y$ for a preparation test $\rho$ is invertible\footnote{In the case of $\reals^+$, this boils down to the usual requirement that the probability be non-vanishing.}, we can \textbf{condition} on $y$ to obtain the state $\frac{1}{\mathbb{P}_{\!\rho}(y)} \rho_y$, which is normalised whenever the original state $\rho$ is:
\begin{equation}\label{conditioningPrepTest}
\begin{tikzpicture}
	\begin{pgfonlayer}{nodelayer}
		\node [style=none] (0) at (-2, -0.25) {};
		\node [style=tightlabelnode] (1) at (1.5, -0.75) {$\frac{1}{\mathbb{P}_{\!\rho}(y)}$};
		\node [style=none] (2) at (-2, 0.25) {};
		\node [style=tightlabelnode] (3) at (2.5, 0.25) {$\SpaceH$};
		\node [style=state] (4) at (-2.5, 0) {$\rho$};
		\node [style=effect] (5) at (0, -0.75) {$y$};
	\end{pgfonlayer}
	\begin{pgfonlayer}{edgelayer}
		\draw [style=dashed, in=180, out=0, looseness=1.50] (0.center) to (5);
		\draw [style=-] (3) to (2.center);
	\end{pgfonlayer}
\end{tikzpicture}
\end{equation}
If $\rho$ is normalised and the $R$-probabilities of all classical outcomes are either zero or invertible, then the reduced state $\restrict{\rho}{\SpaceH}$ can be written as usual in the form of a convex combination of the conditioned states:
\begin{equation}\label{prepTestMarginalConvexCombin}
\begin{tikzpicture}
	\begin{pgfonlayer}{nodelayer}
		\node [style=effect] (0) at (7.5, -0.75) {$y$};
		\node [style=none] (1) at (5.5, -0.25) {};
		\node [style=state] (2) at (5, 0) {$\rho$};
		\node [style=none] (3) at (5.5, 0.25) {};
		\node [style=tightlabelnode] (4) at (10, 0.25) {$\SpaceH$};
		\node [style=tightlabelnode] (5) at (2.5, 0) {$\mathbb{P}_{\!\rho}(y)$};
		\node [style=trace] (6) at (-6.75, -0.75) {};
		\node [style=none] (7) at (-8.75, -0.25) {};
		\node [style=state] (8) at (-9.25, 0) {$\rho$};
		\node [style=none] (9) at (-8.75, 0.25) {};
		\node [style=tightlabelnode] (10) at (-5.5, 0.25) {$\SpaceH$};
		\node [style=tightlabelnode] (11) at (-3.5, 0) {$=$};
		\node [style=tightlabelnode] (12) at (-0.5, -0.25) {$\sum\limits_{y \text{ s.t. } \mathbb{P}_{\!\rho}(y) \neq 0}$};
		\node [style=tightlabelnode] (13) at (9, -0.75) {$\frac{1}{\mathbb{P}_{\!\rho}(y)}$};
		\node [style=tightlabelnode] (14) at (-3.5, -1.5) {};
	\end{pgfonlayer}
	\begin{pgfonlayer}{edgelayer}
		\draw [style=dashed, in=180, out=0, looseness=1.50] (1.center) to (0);
		\draw [style=-] (4) to (3.center);
		\draw [style=dashed, in=180, out=0, looseness=1.50] (7.center) to (6);
		\draw [style=-] (10) to (9.center);
	\end{pgfonlayer}
\end{tikzpicture}
\end{equation}

\subsection{Classical control}

Processes in $R$-probabilistic theories are not limited to tests. In its most general form, a process $f$ includes both some classical output system $Y$ and some \textbf{classical input/control} system $X$:
\begin{equation}\label{generalProcessProbTheory}
\begin{tikzpicture}
	\begin{pgfonlayer}{nodelayer}
		\node [style=box] (0) at (0, 0) {$f$};
		\node [style=tightlabelnode] (1) at (3, -0.75) {$Y$};
		\node [style=none] (2) at (0.5, -0.25) {};
		\node [style=tightlabelnode] (3) at (3, 0.25) {$\SpaceG$};
		\node [style=tightlabelnode] (4) at (-3, 0.25) {$\SpaceH$};
		\node [style=none] (5) at (0.5, 0.25) {};
		\node [style=none] (6) at (-0.5, -0.25) {};
		\node [style=tightlabelnode] (7) at (-3, -0.75) {$X$};
		\node [style=none] (8) at (-0.5, 0.25) {};
	\end{pgfonlayer}
	\begin{pgfonlayer}{edgelayer}
		\draw [style=dashed, in=180, out=0, looseness=1.25] (2.center) to (1);
		\draw [style=-] (3) to (5.center);
		\draw [style=dashed, in=180, out=0, looseness=1.25] (7) to (6.center);
		\draw [style=-] (4) to (8.center);
	\end{pgfonlayer}
\end{tikzpicture}
\end{equation}
The process $f^{(x)}$ corresponding to a definite input value $x \in X$ is immediately obtained by applying $f$ to the deterministic state corresponding to $x$:
\begin{equation}\label{generalProcessControl}
\begin{tikzpicture}
	\begin{pgfonlayer}{nodelayer}
		\node [style=state] (0) at (2.75, -0.75) {$x$};
		\node [style=none] (1) at (5, -0.25) {};
		\node [style=box] (2) at (5.5, 0) {$f$};
		\node [style=tightlabelnode] (3) at (0, 0) {$:=$};
		\node [style=box] (4) at (-4.5, 0) {$f^{(x)}$};
		\node [style=tightlabelnode] (5) at (-7, 0) {$\SpaceH$};
		\node [style=tightlabelnode] (6) at (1.5, 0.25) {$\SpaceH$};
		\node [style=none] (7) at (5, 0.25) {};
		\node [style=tightlabelnode] (8) at (-1.5, 0.25) {$\SpaceG$};
		\node [style=none] (9) at (-4, -0.25) {};
		\node [style=tightlabelnode] (10) at (-1.5, -0.75) {$Y$};
		\node [style=none] (11) at (-4, 0.25) {};
		\node [style=tightlabelnode] (12) at (8.5, 0.25) {$\SpaceG$};
		\node [style=none] (13) at (6, -0.25) {};
		\node [style=tightlabelnode] (14) at (8.5, -0.75) {$Y$};
		\node [style=none] (15) at (6, 0.25) {};
	\end{pgfonlayer}
	\begin{pgfonlayer}{edgelayer}
		\draw [style=dashed, in=0, out=180, looseness=1.75] (1.center) to (0);
		\draw [style=-] (6) to (7.center);
		\draw [style=dashed, in=180, out=0, looseness=1.25] (9.center) to (10);
		\draw [style=-] (8) to (11.center);
		\draw [style=-] (5) to (4);
		\draw [style=dashed, in=180, out=0, looseness=1.25] (13.center) to (14);
		\draw [style=-] (12) to (15.center);
	\end{pgfonlayer}
\end{tikzpicture}
\end{equation}
More generally, we can apply the process $f$ above to an arbitrary state $p:=(p_x)_{x \in X}$ on $X$ to obtain a linear combination of the individual processes $f^{(x)}$ corresponding to each definite input value $x$:
\begin{equation}\label{generalProcessControlLinearComb}
\begin{tikzpicture}
	\begin{pgfonlayer}{nodelayer}
		\node [style=box] (0) at (-4.5, 0) {$f$};
		\node [style=tightlabelnode] (1) at (0, 0) {$=$};
		\node [style=none] (2) at (-5, -0.25) {};
		\node [style=state] (3) at (-7, -0.75) {$p$};
		\node [style=labelnode, inner sep=0.1 mm] (4) at (2, -0.25) {$\sum\limits_{x \in X}$};
		\node [style=none] (5) at (-5, 0.25) {};
		\node [style=tightlabelnode] (6) at (-8, 0.25) {$\SpaceH$};
		\node [style=labelnode] (7) at (3.25, 0) {$p_x$};
		\node [style=tightlabelnode] (8) at (5, 0.25) {$\SpaceH$};
		\node [style=box] (9) at (7.5, 0) {$f^{(x)}$};
		\node [style=none] (10) at (7, 0.25) {};
		\node [style=tightlabelnode] (11) at (-1.5, 0.25) {$\SpaceG$};
		\node [style=none] (12) at (-4, -0.25) {};
		\node [style=tightlabelnode] (13) at (-1.5, -0.75) {$Y$};
		\node [style=none] (14) at (-4, 0.25) {};
		\node [style=tightlabelnode] (15) at (10.5, 0.25) {$\SpaceG$};
		\node [style=none] (16) at (8, -0.25) {};
		\node [style=tightlabelnode] (17) at (10.5, -0.75) {$Y$};
		\node [style=none] (18) at (8, 0.25) {};
	\end{pgfonlayer}
	\begin{pgfonlayer}{edgelayer}
		\draw [style=dashed, in=180, out=0, looseness=1.25] (3) to (2.center);
		\draw [style=-] (6) to (5.center);
		\draw [style=-] (8) to (10.center);
		\draw [style=dashed, in=180, out=0, looseness=1.25] (12.center) to (13);
		\draw [style=-] (11) to (14.center);
		\draw [style=dashed, in=180, out=0, looseness=1.25] (16.center) to (17);
		\draw [style=-] (15) to (18.center);
	\end{pgfonlayer}
\end{tikzpicture}
\end{equation}
When the state $(p_x)_x$ is normalised, this gives a convex combination $\sum_x p_x f^{(x)}$ of the individual processes.

\subsection{Preparations}

As an example of the difference between the constructions above, we look at preparations. In an $R$-probabilistic theory, there are three possible notions of preparation: we have controlled preparations, preparation tests, and convex mixtures of the prepared states. A \textbf{controlled preparation} takes a classical input $x \in X$ and produces a state $f^{(x)}$ in $\SpaceH$, while a preparation test is a state on the joint system $\SpaceH \otimes X$:
\begin{equation}\label{preparationControlledVsTest}
\begin{tikzpicture}
	\begin{pgfonlayer}{nodelayer}
		\node [style=box] (0) at (-6, 0) {$f$};
		\node [style=tightlabelnode] (1) at (-3, 0) {$\SpaceH$};
		\node [style=none] (2) at (-5.5, 0) {};
		\node [style=none] (3) at (-6.5, 0) {};
		\node [style=tightlabelnode] (4) at (-9, 0) {$X$};
		\node [style=none] (5) at (3.5, 0.25) {};
		\node [style=none] (6) at (3.5, -0.25) {};
		\node [style=tightlabelnode] (7) at (6, -0.75) {$X$};
		\node [style=tightlabelnode] (8) at (6, 0.25) {$\SpaceH$};
		\node [style=state] (9) at (3, 0) {$\rho$};
		\node [style=tightlabelnode] (10) at (-6, -2) {controlled preparation};
		\node [style=tightlabelnode] (11) at (4, -2) {preparation test};
	\end{pgfonlayer}
	\begin{pgfonlayer}{edgelayer}
		\draw [style=-] (1) to (2.center);
		\draw [style=dashed, in=180, out=0, looseness=1.25] (4) to (3.center);
		\draw [style=dashed, in=180, out=0, looseness=1.25] (6.center) to (7);
		\draw [style=-] (8) to (5.center);
	\end{pgfonlayer}
\end{tikzpicture}
\end{equation}
Given a normalised controlled preparation and a normalised state $q$ on $X$, i.e. an $R$-distribution $(q_x)_{x \in X}$, we can always obtain a corresponding preparation test as follows:
\begin{equation}\label{preparationTestFromControlledPreparation}
\begin{tikzpicture}
	\begin{pgfonlayer}{nodelayer}
		\node [style=box] (0) at (6, 0.25) {$f$};
		\node [style=tightlabelnode] (1) at (9, 0.25) {$\SpaceH$};
		\node [style=none] (2) at (6.5, 0.25) {};
		\node [style=Zbwdot] (3) at (4, -0.75) {};
		\node [style=state] (4) at (2, -0.75) {$q$};
		\node [style=state] (5) at (-5, 0) {$\rho$};
		\node [style=tightlabelnode] (6) at (-2, -0.75) {$X$};
		\node [style=tightlabelnode] (7) at (-2, 0.25) {$\SpaceH$};
		\node [style=none] (8) at (-4.5, -0.25) {};
		\node [style=none] (9) at (-4.5, 0.25) {};
		\node [style=none] (10) at (5.5, 0.25) {};
		\node [style=none] (11) at (5.5, -1.75) {};
		\node [style=tightlabelnode] (12) at (9, -0.75) {$X$};
		\node [style=tightlabelnode] (13) at (0, 0) {$:=$};
		\node [style=tightlabelnode,text=gray] (14) at (4, 1.75) {classical copy map on $X$};
		\node [style=none] (15) at (4, 1) {};
		\node [style=none] (16) at (4, -0.25) {};
	\end{pgfonlayer}
	\begin{pgfonlayer}{edgelayer}
		\draw [style=-] (1) to (2.center);
		\draw [style=dashed, in=180, out=0, looseness=1.25] (4) to (3);
		\draw [style=dashed, in=180, out=0, looseness=1.25] (8.center) to (6);
		\draw [style=-] (7) to (9.center);
		\draw [style=dashed, in=180, out=0, looseness=1.25] (11.center) to (12);
		\draw [style=dashed, in=180, out=-45] (3) to (11.center);
		\draw [style=dashed, in=180, out=45] (3) to (10.center);
		\draw [style=->, draw=gray] (15.center) to (16.center);
	\end{pgfonlayer}
\end{tikzpicture}
\end{equation}
The normalisation requirement ensures that the $R$-probability of output $x \in X$ in the preparation test is the same specified by the $R$-probability distribution $(q_x)_{x \in X}$:
\begin{equation}\label{preparationTestFromControlledPreparationProb}
\resizebox{\textwidth}{!}{\begin{tikzpicture}
	\begin{pgfonlayer}{nodelayer}
		\node [style=box] (0) at (-5, 0.75) {$f$};
		\node [style=trace] (1) at (-3, 0.75) {};
		\node [style=none] (2) at (-4.5, 0.75) {};
		\node [style=Zbwdot] (3) at (-7, 0) {};
		\node [style=state] (4) at (-9, 0) {$q$};
		\node [style=state] (5) at (-16, 0.5) {$\rho$};
		\node [style=tightlabelnode] (6) at (-13, -0.75) {$X$};
		\node [style=trace] (7) at (-13.5, 0.75) {};
		\node [style=none] (8) at (-15.5, 0.25) {};
		\node [style=none] (9) at (-15.5, 0.75) {};
		\node [style=none] (10) at (-5.5, 0.75) {};
		\node [style=none] (11) at (-5.5, -0.75) {};
		\node [style=tightlabelnode] (12) at (-2, -0.75) {$X$};
		\node [style=tightlabelnode] (13) at (-11, 0) {$=$};
		\node [style=tightlabelnode] (14) at (0, 0) {$=$};
		\node [style=state] (15) at (2, 0) {$q$};
		\node [style=tightlabelnode] (16) at (6.5, -0.75) {$X$};
		\node [style=tightlabelnode] (17) at (8.5, 0) {$=$};
		\node [style=Zbwdot] (18) at (4, 0) {};
		\node [style=trace] (19) at (5.5, 0.75) {};
		\node [style=none] (20) at (5.5, -0.75) {};
		\node [style=state] (21) at (10.5, 0) {$q$};
		\node [style=tightlabelnode] (22) at (13, 0) {$X$};
	\end{pgfonlayer}
	\begin{pgfonlayer}{edgelayer}
		\draw [style=-] (1) to (2.center);
		\draw [style=dashed, in=180, out=0, looseness=1.25] (4) to (3);
		\draw [style=dashed, in=180, out=0, looseness=1.25] (8.center) to (6);
		\draw [style=-] (7) to (9.center);
		\draw [style=dashed, in=180, out=0, looseness=1.25] (11.center) to (12);
		\draw [style=dashed, in=180, out=-45] (3) to (11.center);
		\draw [style=dashed, in=180, out=45] (3) to (10.center);
		\draw [style=dashed, in=180, out=0, looseness=1.25] (15) to (18);
		\draw [style=dashed, in=180, out=0, looseness=1.25] (20.center) to (16);
		\draw [style=dashed, in=180, out=-45] (18) to (20.center);
		\draw [style=dashed, in=180, out=45] (18) to (19);
		\draw [style=dashed, in=180, out=0, looseness=1.25] (21) to (22);
	\end{pgfonlayer}
\end{tikzpicture}}
\end{equation}
Given a preparation test obtained as in Equation \ref{preparationTestFromControlledPreparation}, one can further discard the classical output system $X$ to obtain the corresponding convex combination of prepared states, as shown in Equation \ref{prepTestMarginalConvexCombin}. There is no compositional way of obtaining the controlled preparation back from the preparation test, or the preparation test back from the convex combination of prepared states.

\subsection{Coarse-graining}

Sometimes the classical outputs of a process $f$  are not immediately interesting on their own, and one might wish to apply some form of \textbf{classical post-processing} to them, by which we mean the application of some additional classical (probabilistic) process to the output system. Most commonly, one obtains a new process $g$ by applying a deterministic function $q : X \rightarrow Z$ to the relevant output system $X$ of $f$, mapping the original outputs to some \inlineQuote{more interesting} values in some other set $Z$:
\begin{equation}\label{classicalPostProcessing}
\begin{tikzpicture}
	\begin{pgfonlayer}{nodelayer}
		\node [style=box] (0) at (1.5, 0) {$f$};
		\node [style=tightlabelnode] (1) at (4, -0.75) {$Y$};
		\node [style=none] (2) at (2, -0.25) {};
		\node [style=tightlabelnode] (3) at (-1, 0.25) {$\SpaceH$};
		\node [style=none] (4) at (2, 0.25) {};
		\node [style=none] (5) at (1, -0.25) {};
		\node [style=tightlabelnode] (6) at (-1, -0.75) {$X$};
		\node [style=none] (7) at (1, 0.25) {};
		\node [style=box] (8) at (6, -0.75) {$q$};
		\node [style=tightlabelnode] (9) at (8, 0.25) {$\SpaceG$};
		\node [style=tightlabelnode] (10) at (8, -0.75) {$Z$};
		\node [style=tightlabelnode] (11) at (-3, 0) {$:=$};
		\node [style=none] (12) at (-8, 0.25) {};
		\node [style=none] (13) at (-7, -0.25) {};
		\node [style=tightlabelnode] (14) at (-5, -0.75) {$Z$};
		\node [style=tightlabelnode] (15) at (-10, 0.25) {$\SpaceH$};
		\node [style=box] (16) at (-7.5, 0) {$g$};
		\node [style=tightlabelnode] (17) at (-5, 0.25) {$\SpaceG$};
		\node [style=none] (18) at (-7, 0.25) {};
		\node [style=none] (19) at (-8, -0.25) {};
		\node [style=tightlabelnode] (20) at (-10, -0.75) {$X$};
	\end{pgfonlayer}
	\begin{pgfonlayer}{edgelayer}
		\draw [style=dashed, in=180, out=0, looseness=1.25] (2.center) to (1);
		\draw [style=dashed, in=180, out=0, looseness=1.25] (6) to (5.center);
		\draw [style=-] (3) to (7.center);
		\draw [style=dashed] (1) to (8);
		\draw [style=dashed] (8) to (10);
		\draw [style=-] (4.center) to (9);
		\draw [style=dashed, in=180, out=0, looseness=1.25] (13.center) to (14);
		\draw [style=dashed, in=180, out=0, looseness=1.25] (20) to (19.center);
		\draw [style=-] (15) to (12.center);
		\draw [style=-] (18.center) to (17);
	\end{pgfonlayer}
\end{tikzpicture}
\end{equation}
If we now focus on our new values in $Z$, we see that the process we obtain is exactly what the OPT framework refers to as a \textit{coarse-graining} of process $f$:
\begin{equation}\label{coarseGraining}
\begin{tikzpicture}
	\begin{pgfonlayer}{nodelayer}
		\node [style=box] (0) at (-2.75, 0) {$f$};
		\node [style=tightlabelnode] (1) at (-0.5, -0.75) {$Y$};
		\node [style=none] (2) at (-2.25, -0.25) {};
		\node [style=tightlabelnode] (3) at (-5, 0.25) {$\SpaceH$};
		\node [style=none] (4) at (-2.25, 0.25) {};
		\node [style=none] (5) at (-3.25, -0.25) {};
		\node [style=tightlabelnode] (6) at (-5, -0.75) {$X$};
		\node [style=none] (7) at (-3.25, 0.25) {};
		\node [style=box] (8) at (1.25, -0.75) {$q$};
		\node [style=tightlabelnode] (9) at (4.25, 0.25) {$\SpaceG$};
		\node [style=effect] (10) at (3.25, -0.75) {$z$};
		\node [style=tightlabelnode] (11) at (-7, 0) {$=$};
		\node [style=none] (12) at (-12, 0.25) {};
		\node [style=tightlabelnode] (13) at (-14, 0.25) {$\SpaceH$};
		\node [style=box] (14) at (-11.5, 0) {$g_z$};
		\node [style=tightlabelnode] (15) at (-9, 0) {$\SpaceG$};
		\node [style=none] (16) at (-11, 0) {};
		\node [style=none] (17) at (-12, -0.25) {};
		\node [style=tightlabelnode] (18) at (-14, -0.75) {$X$};
		\node [style=tightlabelnode] (19) at (6, 0) {$=$};
		\node [style=tightlabelnode] (20) at (8.25, -0.25) {$\sum\limits_{y \text{ s.t. } q(y) = z}$};
		\node [style=none] (21) at (13.75, 0) {};
		\node [style=none] (22) at (12.75, 0.25) {};
		\node [style=tightlabelnode] (23) at (15.75, 0) {$\SpaceG$};
		\node [style=tightlabelnode] (24) at (10.75, -0.75) {$X$};
		\node [style=tightlabelnode] (25) at (10.75, 0.25) {$\SpaceH$};
		\node [style=none] (26) at (12.75, -0.25) {};
		\node [style=box] (27) at (13.25, 0) {$f_y$};
	\end{pgfonlayer}
	\begin{pgfonlayer}{edgelayer}
		\draw [style=dashed, in=180, out=0, looseness=1.25] (2.center) to (1);
		\draw [style=dashed, in=180, out=0, looseness=1.25] (6) to (5.center);
		\draw [style=-] (3) to (7.center);
		\draw [style=dashed] (1) to (8);
		\draw [style=dashed] (8) to (10);
		\draw [style=-, in=180, out=0] (4.center) to (9);
		\draw [style=dashed, in=180, out=0, looseness=1.25] (18) to (17.center);
		\draw [style=-] (13) to (12.center);
		\draw [style=-] (16.center) to (15);
		\draw [style=dashed, in=180, out=0, looseness=1.25] (24) to (26.center);
		\draw [style=-] (25) to (22.center);
		\draw [style=-] (21.center) to (23);
	\end{pgfonlayer}
\end{tikzpicture}
\end{equation}
As a consequence, by a \textbf{coarse-graining} of a process $f$ we will mean a process $g$ taking the form of Equation \ref{classicalPostProcessing} for some deterministic function $q: X \rightarrow Z$. Just like in the OPT framework, we say that $f$ is a \textbf{refinement} of $g$ whenever $g$ is a coarse-graining of $f$.

\subsection{Sharp preparations/observation pairs}

A controlled preparation can be seen as a way of encoding the values of its classical input set $X$ into its output system $\SpaceH$. However, not all preparations admit a corresponding decoding process which deterministically returns the original input values\footnote{The simplest example being given by non-injective deterministic functions.}: we refer to those preparations admitting one such decoding as \textbf{sharp preparations}, and they correspond to perfectly distinguishable states \cite{Chiribella2010}. We say that a pair $(p,m)$ of a controlled preparation $p: X \rightarrow \SpaceH$ and an observation test $m:\SpaceH \rightarrow X$ is a \textbf{sharp preparation/observation pair} (or \textbf{SPO pair}, for short) if the observation test $m$ witnesses the sharpness of the controlled preparation $p$:
\begin{equation}\label{sharpPreparationObservationPair}
\begin{tikzpicture}
	\begin{pgfonlayer}{nodelayer}
		\node [style=tightlabelnode] (0) at (-5, -1.5) {$X$};
		\node [style=box] (1) at (-7, -1.5) {$m$};
		\node [style=tightlabelnode] (2) at (-9, -1.5) {$\SpaceH$};
		\node [style=box] (3) at (-7, 1.5) {$p$};
		\node [style=tightlabelnode] (4) at (-9, 1.5) {$X$};
		\node [style=tightlabelnode] (5) at (-5, 1.5) {$\SpaceH$};
		\node [style=box] (6) at (3.5, 0) {$m$};
		\node [style=tightlabelnode] (7) at (-1, 0) {$X$};
		\node [style=box] (8) at (1, 0) {$p$};
		\node [style=tightlabelnode] (9) at (5.5, 0) {$X$};
		\node [style=tightlabelnode] (10) at (7, 0) {$=$};
		\node [style=tightlabelnode] (11) at (8.5, 0) {$X$};
		\node [style=tightlabelnode] (12) at (10.5, 0) {$X$};
		\node [style=tightlabelnode] (13) at (-7, -2.75) {observation};
		\node [style=tightlabelnode] (14) at (-7, 0.25) {preparation};
	\end{pgfonlayer}
	\begin{pgfonlayer}{edgelayer}
		\draw [style=-] (2) to (1);
		\draw [style=-] (3) to (5);
		\draw [style=-] (8) to (6);
		\draw [style=dashed] (1) to (0);
		\draw [style=dashed] (4) to (3);
		\draw [style=dashed] (7) to (8);
		\draw [style=dashed] (6) to (9);
		\draw [style=dashed] (11) to (12);
	\end{pgfonlayer}
\end{tikzpicture}
\end{equation}
We say that an SPO pair is \textbf{normalised}\footnote{To be precise, it is normalised SPOs that correspond to perfectly distinguishable states in OPTs.} when both $p$ and $m$ are normalised processes. In fact, it suffices to ask for $m$ to be normalised, as normalisation of $p$ always follows\footnote{The converse is not in general true. As a simple counterexample, take a deterministic function $p: X \rightarrow Y$ which is injective but not surjective, and let $m: Y \rightharpoonup X$ be its left inverse, which is never total: then $p$ is normalised, as it is a total function, but $m$ is not total, and hence not normalised.}:  
\begin{equation}\label{sharpPreparationObservationPairNormalisation}
\begin{tikzpicture}
	\begin{pgfonlayer}{nodelayer}
		\node [style=box] (0) at (2, 0) {$m$};
		\node [style=tightlabelnode] (1) at (-2.5, 0) {};
		\node [style=box] (2) at (-0.5, 0) {$p$};
		\node [style=trace] (3) at (4, 0) {};
		\node [style=tightlabelnode] (4) at (5.5, 0) {$=$};
		\node [style=tightlabelnode] (5) at (7, 0) {};
		\node [style=trace] (6) at (9, 0) {};
		\node [style=box] (7) at (-8, 0) {$p$};
		\node [style=trace] (8) at (-5.5, 0) {};
		\node [style=tightlabelnode] (9) at (-10, 0) {};
		\node [style=tightlabelnode] (10) at (-4, 0) {$=$};
	\end{pgfonlayer}
	\begin{pgfonlayer}{edgelayer}
		\draw [style=-] (2) to (0);
		\draw [style=dashed] (1) to (2);
		\draw [style=dashed] (0) to (3);
		\draw [style=dashed] (5) to (6);
		\draw [style=-] (7) to (8);
		\draw [style=dashed] (9) to (7);
	\end{pgfonlayer}
\end{tikzpicture}
\end{equation}

\subsection{Decoherence maps}

If $(p,m)$ is a normalised SPO pair, we refer to the following process as the associated \textbf{decoherence map}:
\begin{equation}\label{decoherenceMap}
\begin{tikzpicture}
	\begin{pgfonlayer}{nodelayer}
		\node [style=box] (0) at (-4.5, 0) {$\decoh{p,m}$};
		\node [style=tightlabelnode] (1) at (-7.5, 0) {$\SpaceH$};
		\node [style=tightlabelnode] (2) at (-1.5, 0) {$\SpaceH$};
		\node [style=box] (3) at (3.5, 0) {$m$};
		\node [style=tightlabelnode] (4) at (1.5, 0) {$\SpaceH$};
		\node [style=box] (5) at (6, 0) {$p$};
		\node [style=tightlabelnode] (6) at (8, 0) {$\SpaceH$};
		\node [style=tightlabelnode] (7) at (-4.5, -1.25) {decoherence};
		\node [style=tightlabelnode] (8) at (0, 0) {$:=$};
	\end{pgfonlayer}
	\begin{pgfonlayer}{edgelayer}
		\draw [style=-] (0) to (2);
		\draw [style=-] (1) to (0);
		\draw [style=dashed] (3) to (5);
		\draw [style=-] (4) to (3);
		\draw [style=-] (5) to (6);
	\end{pgfonlayer}
\end{tikzpicture}
\end{equation} 
In order to keep track of the classical system $X$ intervening in the decoherence map, we will use $\classicalSystem{p,m}$ to denote it. There is good reason for using the name \inlineQuote{decoherence map} in this context. Firstly, the process $\decoh{p,m}$ is normalised (because composition of normalised processes) and idempotent:
\begin{equation}\label{decoherenceMapIdemp}
\resizebox{\textwidth}{!}{\begin{tikzpicture}
	\begin{pgfonlayer}{nodelayer}
		\node [style=box] (0) at (16, 0) {$\decoh{p,m}$};
		\node [style=tightlabelnode] (1) at (13.5, 0) {};
		\node [style=tightlabelnode] (2) at (18.5, 0) {};
		\node [style=box] (3) at (-12.5, 0) {$m$};
		\node [style=tightlabelnode] (4) at (-14.5, 0) {};
		\node [style=box] (5) at (-9.5, 0) {$p$};
		\node [style=tightlabelnode] (6) at (-16, 0) {$=$};
		\node [style=tightlabelnode] (7) at (0, 0) {$=$};
		\node [style=tightlabelnode] (8) at (1.5, 0) {};
		\node [style=tightlabelnode] (9) at (10.5, 0) {};
		\node [style=box] (10) at (3.5, 0) {$m$};
		\node [style=box] (11) at (8.5, 0) {$p$};
		\node [style=tightlabelnode] (12) at (12, 0) {$=$};
		\node [style=tightlabelnode] (13) at (-1.5, 0) {};
		\node [style=box] (14) at (-6.5, 0) {$m$};
		\node [style=box] (15) at (-3.5, 0) {$p$};
		\node [style=tightlabelnode] (16) at (-25.5, 0) {};
		\node [style=box] (17) at (-23, 0) {$\decoh{p,m}$};
		\node [style=tightlabelnode] (18) at (-17.5, 0) {};
		\node [style=box] (19) at (-20, 0) {$\decoh{p,m}$};
	\end{pgfonlayer}
	\begin{pgfonlayer}{edgelayer}
		\draw [style=-] (0) to (2);
		\draw [style=-] (1) to (0);
		\draw [style=dashed] (3) to (5);
		\draw [style=-] (4) to (3);
		\draw [style=dashed] (10) to (11);
		\draw [style=-] (8) to (10);
		\draw [style=-] (11) to (9);
		\draw [style=dashed] (14) to (15);
		\draw [style=-] (15) to (13);
		\draw [style=-] (5) to (14);
		\draw [style=-] (16) to (17);
		\draw [style=-] (19) to (18);
		\draw [style=-] (17) to (19);
	\end{pgfonlayer}
\end{tikzpicture}}
\end{equation} 
Secondly, decoherence maps can be used to reconstruct classical systems from arbitrary ones: to understand how this works, we use a mild variation on a common construction from category theory, known as \textit{Karoubi envelope} (or \textit{idempotent completion}). If $\CategoryC$ is a SMC with an environment structure, then we define the \textbf{normalised Karoubi envelope} of $\CategoryC$, denoted by $\normalisedKaroubiEnvelope{\CategoryC}$, to be the SMC defined as follows:
\begin{itemize}
	\item the systems are given by  pairs in the form $(\SpaceH, h)$, where $\SpaceH$ is a system of $\CategoryC$ and $h: \SpaceH \rightarrow \SpaceH$ is a normalised idempotent process;
	\item the processes $f: (\SpaceH,h) \rightarrow (\SpaceG,g)$ are exactly those processes $f: \SpaceH \rightarrow \SpaceG$ in $\CategoryC$ which are invariant under the specified idempotents, i.e. those which satisfy $f = g \circ f \circ h$;
	\item the tensor product on systems is given by $(\SpaceH,h) \otimes (\SpaceG,g) := (\SpaceH \otimes \SpaceG, h \otimes g)$, while the tensor product on processes is the one inherited from $\CategoryC$.
\end{itemize}
By looking at objects in the form $(\SpaceH, \id{\SpaceH})$, it is immediate to see that $\CategoryC$ is a full sub-SMC of $\normalisedKaroubiEnvelope{\CategoryC}$; in particular, the two categories have the same scalars. In fact, we can prove that $\normalisedKaroubiEnvelope{\CategoryC}$ is an $R$-probabilistic theory whenever $\CategoryC$ is.
\begin{lemma}\label{lem_SplitProbabilisticTheory}
If $\CategoryC$ is an $R$-probabilistic theory, then so is $\normalisedKaroubiEnvelope{\CategoryC}$, with classical systems, enrichment and discarding maps directly inherited from $\CategoryC$. 
\end{lemma}

\noindent Amongst the many objects of the normalised Karoubi envelope for an $R$-probabilistic theory sit all objects in the form $(\SpaceH,\decoh{p,m})$, with $(p,m)$ a normalised SPO pair on $\SpaceH$: we refer to systems in this form as \textbf{decohered systems}, and we now show that they can be used to recover the classical systems of the theory.
\begin{theorem}[\textbf{Decohered systems}]\label{thm_decoheredSystems}\hfill\\
Let $\CategoryC$ be an $R$-probabilistic theory, and let $\decoheredSystemsSubcategory{\CategoryC}$ be the full subcategory of $\normalisedKaroubiEnvelope{\CategoryC}$ spanned by the decohered systems. Then $\decoheredSystemsSubcategory{\CategoryC}$ is a sub-SMC, and it is equivalent to the sub-SMC $\classicalSubcategory{\CategoryC}$ of classical systems in $\CategoryC$. The equivalence sends a decohered system $(\SpaceH, \decoh{p,m})$ to the classical system $\classicalSystem{p,m}$, and a process $f: (\SpaceH, \decoh{p,m}) \rightarrow (\SpaceG, \decoh{q,n})$ to the following classical process:
\begin{equation}\label{decoheredProcessesClassical}
\begin{tikzpicture}
	\begin{pgfonlayer}{nodelayer}
		\node [style=box] (0) at (-2.5, 0) {$p$};
		\node [style=tightlabelnode] (1) at (5.5, 0) {$\classicalSystem{q,n}$};
		\node [style=tightlabelnode] (2) at (-5.5, 0) {$\classicalSystem{p,m}$};
		\node [style=box] (3) at (2.5, 0) {$n$};
		\node [style=box] (4) at (0, 0) {$f$};
	\end{pgfonlayer}
	\begin{pgfonlayer}{edgelayer}
		\draw [style=dashed] (3) to (1);
		\draw [style=dashed] (2) to (0);
		\draw [style=-] (0) to (4);
		\draw [style=-] (4) to (3);
	\end{pgfonlayer}
\end{tikzpicture}
\end{equation} 
\end{theorem}

In the case of quantum theory, the decoherence maps associated with a (possibly degenerate) observable arise from the normalised SPO pair given by considering the demolition measurement in the observable, together with its adjoint. However, there are more decoherence maps in quantum theory than those associated with observables. Indeed, we have imposed no requirement for the observation to be adjoint to the preparation: hence decoherence maps are always normalised and idempotent, but not necessarily self-adjoint. As such, our notion of decoherence maps in quantum theory is more general than the self-adjoint normalised idempotent notion used in Ref. \cite{Selinger2008} and related works. 

\newpage
\section{Quantum Theory}
\label{section_QT}

In CQM, dagger compact categories are taken to be abstract models of pure-state quantum theory, and the corresponding models of mixed-state quantum theory are obtained via the CPM construction \cite{Selinger2007}. The CPM category $\CPMCategory{\fdHilbCategory}$ contains all positive states and completely positive maps of finite-dimensional quantum theory, it comes with an environment structure given by the trace, and it can be enriched in commutative monoids, with $\reals^+$ as semiring of scalars (the usual $\reals^+$-linear structure of CP maps). Unfortunately, $\CPMCategory{\fdHilbCategory}$ does not come with a way of talking about classical systems, and in order to do so one usually appeals to the CP* construction \cite{Coecke2014a,Cunningham2015}, which is related to C* algebras and quantum logic. In this work, we will not appeal to the CP* construction: instead, we will rely on the normalised Karoubi envelope to get a probabilistic theory out of $\CPMCategory{\fdHilbCategory}$, following the steps of \cite{Selinger2008} (a similar approach is followed by the recent \cite{Tull2017}).

As we have seen before, the normalised Karoubi envelope $\normalisedKaroubiEnvelope{\CPMCategory{\fdHilbCategory}}$ contains $\CPMCategory{\fdHilbCategory}$ as a full sub-SMC, and we will refer to the objects in that sub-SMC as \textbf{quantum systems}. It also contains objects in the form $(\SpaceH,\decoh{\hbox{\begin{tikzpicture} [scale=1.2,transform shape] 

\def\deltax{0.3} 
\def\deltay{0.5} 


\node [dot, fill=\Zbwcolour] (mult) at (0,0) {};

\end{tikzpicture}
}\!\!})$, where $\hbox{\begin{tikzpicture} [scale=1.2,transform shape] 

\def\deltax{0.3} 
\def\deltay{0.5} 


\node [dot, fill=\Zbwcolour] (mult) at (0,0) {};

\end{tikzpicture}
}\!\!$ is a special commutative $\dagger$-Frobenius algebra on the finite-dimensional Hilbert space $\SpaceH$: we will refer to objects in this form as \textbf{classical systems}. The \textbf{decoherence map} $\decoh{\hbox{\begin{tikzpicture} [scale=1.2,transform shape] 

\def\deltax{0.3} 
\def\deltay{0.5} 


\node [dot, fill=\Zbwcolour] (mult) at (0,0) {};

\end{tikzpicture}
}\!\!}$ for $\hbox{\begin{tikzpicture} [scale=1.2,transform shape] 

\def\deltax{0.3} 
\def\deltay{0.5} 


\node [dot, fill=\Zbwcolour] (mult) at (0,0) {};

\end{tikzpicture}
}\!\!$ is the normalised idempotent process defined as follows: 
\begin{equation}\label{decoherenceCPMexplicit}
\begin{tikzpicture}
	\begin{pgfonlayer}{nodelayer}
		\node [style=box] (0) at (-9, 0) {$\decoh{\hbox{\input{symbols/ZbwdotSym.tex}}\!\!}$};
		\node [style=tightlabelnode] (1) at (-12, 0) {};
		\node [style=tightlabelnode] (2) at (-6, 0) {};
		\node [style=tightlabelnode] (3) at (-4, 0) {$:=$};
		\node [style=tightlabelnode] (4) at (-2, 0) {};
		\node [style=Zbwdot] (5) at (0, 0) {};
		\node [style=none] (6) at (1.25, -0.75) {};
		\node [style=trace] (7) at (1.25, 0.75) {};
		\node [style=tightlabelnode] (8) at (2.5, -0.75) {};
		\node [style=tightlabelnode] (9) at (15, 0) {};
		\node [style=state] (10) at (13, 0) {$x$};
		\node [style=tightlabelnode] (11) at (4.5, 0) {$=$};
		\node [style=effect] (12) at (11, 0) {$x$};
		\node [style=tightlabelnode] (13) at (9, 0) {};
		\node [style=tightlabelnode] (14) at (7, -0.25) {$\sum\limits_{x \in \classicalStates{\hbox{\input{symbols/ZbwdotSym.tex}}\!\!}}$};
	\end{pgfonlayer}
	\begin{pgfonlayer}{edgelayer}
		\draw [style=-] (0) to (2);
		\draw [style=-] (1) to (0);
		\draw [style=-, in=180, out=-45] (5) to (6.center);
		\draw [style=-, in=0, out=180] (5) to (4);
		\draw [style=-, in=180, out=45] (5) to (7);
		\draw [style=-] (6.center) to (8);
		\draw [style=-, in=0, out=180] (12) to (13);
		\draw [style=-] (10) to (9);
	\end{pgfonlayer}
\end{tikzpicture}
\end{equation}  
where $(\ket{x})_{x \in \classicalStates{\hbox{\begin{tikzpicture} [scale=1.2,transform shape] 

\def\deltax{0.3} 
\def\deltay{0.5} 


\node [dot, fill=\Zbwcolour] (mult) at (0,0) {};

\end{tikzpicture}
}\!\!}}$ is the orthonormal basis formed by the classical states of $\hbox{\begin{tikzpicture} [scale=1.2,transform shape] 

\def\deltax{0.3} 
\def\deltay{0.5} 


\node [dot, fill=\Zbwcolour] (mult) at (0,0) {};

\end{tikzpicture}
}\!\!$ (see Ref. \cite{Coecke2013b}).
\begin{theorem}[\textbf{Classical systems} \cite{Selinger2008}]\label{thm_classicalSystemsCPM}\hfill\\
Let $\CategoryD$ be the full subcategory of $\normalisedKaroubiEnvelope{\CPMCategory{\fdHilbCategory}}$ spanned by the classical systems. Then $\CategoryD$ is a full sub-SMC equivalent to $\RMatCategory{\reals^+}$. The equivalence sends a classical system $(\SpaceH, \decoh{\hbox{\begin{tikzpicture} [scale=1.2,transform shape] 

\def\deltax{0.3} 
\def\deltay{0.5} 


\node [dot, fill=\Zbwcolour] (mult) at (0,0) {};

\end{tikzpicture}
}\!\!})$ to the set $\classicalStates{\hbox{\begin{tikzpicture} [scale=1.2,transform shape] 

\def\deltax{0.3} 
\def\deltay{0.5} 


\node [dot, fill=\Zbwcolour] (mult) at (0,0) {};

\end{tikzpicture}
}\!\!}$ of classical states for $\hbox{\begin{tikzpicture} [scale=1.2,transform shape] 

\def\deltax{0.3} 
\def\deltay{0.5} 


\node [dot, fill=\Zbwcolour] (mult) at (0,0) {};

\end{tikzpicture}
}\!\!$, and a process $f: (\SpaceH,\decoh{\hbox{\begin{tikzpicture} [scale=1.2,transform shape] 

\def\deltax{0.3} 
\def\deltay{0.5} 


\node [dot, fill=\Zbwcolour] (mult) at (0,0) {};

\end{tikzpicture}
}\!\!}) \rightarrow (\SpaceG, \decoh{\hbox{\begin{tikzpicture} [scale=1.2,transform shape] 

\def\deltax{0.3} 
\def\deltay{0.5} 


\node [dot, fill=\Xbwcolour] (mult) at (0,0) {};

\end{tikzpicture}
}\!\!})$ to the following classical process:
\begin{equation}\label{decoheredProcessesClassicalCPM}
\Big( \bra{y}f\ket{x} \Big)_{(y,x) \in \classicalStates{\hbox{\begin{tikzpicture} [scale=1.2,transform shape] 

\def\deltax{0.3} 
\def\deltay{0.5} 


\node [dot, fill=\Xbwcolour] (mult) at (0,0) {};

\end{tikzpicture}
}\!\!} \times \classicalStates{\hbox{\begin{tikzpicture} [scale=1.2,transform shape] 

\def\deltax{0.3} 
\def\deltay{0.5} 


\node [dot, fill=\Zbwcolour] (mult) at (0,0) {};

\end{tikzpicture}
}\!\!}} : \classicalStates{\hbox{\begin{tikzpicture} [scale=1.2,transform shape] 

\def\deltax{0.3} 
\def\deltay{0.5} 


\node [dot, fill=\Zbwcolour] (mult) at (0,0) {};

\end{tikzpicture}
}\!\!} \rightarrow \classicalStates{\hbox{\begin{tikzpicture} [scale=1.2,transform shape] 

\def\deltax{0.3} 
\def\deltay{0.5} 


\node [dot, fill=\Xbwcolour] (mult) at (0,0) {};

\end{tikzpicture}
}\!\!}
\end{equation} 
\end{theorem}
\noindent When talking about \textbf{quantum theory} in the context of probabilistic theories, we will refer to the full sub-SMC of $\normalisedKaroubiEnvelope{\CPMCategory{\fdHilbCategory}}$ jointly spanned by the quantum systems and the classical systems.

\section{Causality and no-signalling}
\label{section_nosig}


Causality is perhaps the trickiest point of contact between the OPT framework and probabilistic theories as we have defined them. On the one hand, causality is part of the very definition of probabilistic theories: a distinguished family of discarding maps is singled out by the basic operational requirement that states be localisable. This is how causality is usually understood in the CQM community \cite{Coecke2013a,Coecke2016}. 
On the other hand, however, proving causality in a purely compositional way is impossible without reference to that very environment structure. Indeed, one encounters the following issue when trying to formulate the causality axiom as understood by the OPT community: the observation tests in OPTs have additional requirements that make them \inlineQuote{normalised} in a suitable sense, while the ones defined in this work have no such requirement placed upon them. In order to single out a set of observation tests which would be suitable to formulate the causality axiom (as understood in OPTs), we would have to somehow refer to the environment structure to impose an appropriate normalisation condition\footnote{For example, note that $\CPMCategory{\fdHilbCategory}$ admits environment structures different from the canonical one.}. 

As a consequence of the remarks above, we will not talk about a \textit{causality axiom} in the context of probabilistic theories. Instead, we will keep in mind the following general scenario, involving a (normalised) test $f$ followed by a normalised process $g$:
\begin{equation}\label{causalityScenario}
\begin{tikzpicture}
	\begin{pgfonlayer}{nodelayer}
		\node [style=box] (0) at (-1.5, 0.25) {$f$};
		\node [style=none] (1) at (1.5, 1) {};
		\node [style=none] (2) at (-1, 0.5) {};
		\node [style=none] (3) at (1, 0) {};
		\node [style=tightlabelnode] (4) at (-4, 0.25) {};
		\node [style=none] (5) at (-1, 0) {};
		\node [style=none] (6) at (-2, 0.25) {};
		\node [style=none] (7) at (-0.5, -1) {};
		\node [style=none] (8) at (1, -0.5) {};
		\node [style=none] (9) at (2, -0.25) {};
		\node [style=tightlabelnode] (10) at (4, -0.25) {};
		\node [style=box] (11) at (1.5, -0.25) {$g$};
		\node [style=tightlabelnode] (12) at (4, 1) {};
		\node [style=tightlabelnode] (13) at (-4, -1) {};
	\end{pgfonlayer}
	\begin{pgfonlayer}{edgelayer}
		\draw [style=dashed, in=180, out=0, looseness=1.25] (2.center) to (1.center);
		\draw [style=-] (3.center) to (5.center);
		\draw [style=-] (4) to (6.center);
		\draw [style=-] (10) to (9.center);
		\draw [style=dashed, in=180, out=0, looseness=1.25] (7.center) to (8.center);
		\draw [style=dashed] (7.center) to (13);
		\draw [style=dashed] (1.center) to (12);
	\end{pgfonlayer}
\end{tikzpicture}
\end{equation} 
By \textbf{causality} we will mean the \textit{observation} that the local output of the test $f$ is independent of the input of the following process $g$, i.e. that normalised processes satisfy \inlineQuote{no-signalling from the future} \cite{Chiribella2010}:
\begin{equation}\label{causalityScenarioNoSig}
\begin{tikzpicture}
	\begin{pgfonlayer}{nodelayer}
		\node [style=box] (0) at (-7.5, 0) {$f$};
		\node [style=none] (1) at (-4.5, 0.75) {};
		\node [style=none] (2) at (-7, 0.25) {};
		\node [style=none] (3) at (-5, -0.25) {};
		\node [style=tightlabelnode] (4) at (-10, 0) {};
		\node [style=none] (5) at (-7, -0.25) {};
		\node [style=none] (6) at (-8, 0) {};
		\node [style=none] (7) at (-6.5, -1.25) {};
		\node [style=none] (8) at (-5, -0.75) {};
		\node [style=none] (9) at (-4, -0.5) {};
		\node [style=trace] (10) at (-2.5, -0.5) {};
		\node [style=box] (11) at (-4.5, -0.5) {$g$};
		\node [style=tightlabelnode] (12) at (-2, 0.75) {};
		\node [style=tightlabelnode] (13) at (-10, -1.25) {};
		\node [style=tightlabelnode] (14) at (0, 0) {$=$};
		\node [style=tightlabelnode] (15) at (2, -1.25) {};
		\node [style=none] (16) at (4, 0) {};
		\node [style=tightlabelnode] (17) at (8, 0.75) {};
		\node [style=tightlabelnode] (18) at (2, 0) {};
		\node [style=box] (19) at (4.5, 0) {$f$};
		\node [style=none] (20) at (5, -0.25) {};
		\node [style=trace] (21) at (7, -0.25) {};
		\node [style=none] (22) at (5, 0.25) {};
		\node [style=none] (23) at (6.5, 0.75) {};
		\node [style=trace] (24) at (6, -1.25) {};
	\end{pgfonlayer}
	\begin{pgfonlayer}{edgelayer}
		\draw [style=dashed, in=180, out=0, looseness=1.25] (2.center) to (1.center);
		\draw [style=-] (3.center) to (5.center);
		\draw [style=-] (4) to (6.center);
		\draw [style=-] (10) to (9.center);
		\draw [style=dashed, in=180, out=0, looseness=1.25] (7.center) to (8.center);
		\draw [style=dashed] (7.center) to (13);
		\draw [style=dashed] (1.center) to (12);
		\draw [style=dashed, in=180, out=0, looseness=1.25] (22.center) to (23.center);
		\draw [style=-] (21) to (20.center);
		\draw [style=-] (18) to (16.center);
		\draw [style=dashed] (24) to (15);
		\draw [style=dashed] (23.center) to (17);
	\end{pgfonlayer}
\end{tikzpicture}
\end{equation}


More generally, when talking about \textbf{no-signalling} we have in mind some $N$-party \textbf{Bell-type measurement scenario} \cite{Bell,Pusey}, which we define to be a process in the following form (where the processes $B_1,...,B_N$ and the state $\rho$ involved in constructing the scenario are all normalised):
\begin{equation}\label{BellTest}
\begin{tikzpicture}
	\begin{pgfonlayer}{nodelayer}
		\node [style=none] (0) at (-2, 1) {};
		\node [style=none] (1) at (-2, -1) {};
		\node [style=none, doubled] (2) at (0.5, 2.5) {};
		\node [style=none] (3) at (2.5, -2.5) {};
		\node [style=none, doubled] (4) at (0.5, -2.5) {};
		\node [style=none] (5) at (2.5, 2.5) {};
		\node [style=box] (6) at (0, 2.5) {$B_1$};
		\node [style=box] (7) at (0, -2.5) {$B_{N}$};
		\node [style=tightlabelnode] (8) at (0, 0) {$\goodvdots$};
		\node [style=tightlabelnode] (9) at (3.25, 2.5) {\small{$O_1$}};
		\node [style=tightlabelnode] (10) at (3.25, -2.5) {\small{$O_N$}};
		\node [style=tightlabelnode] (11) at (3.25, 0) {$\goodvdots$};
		\node [style=none] (12) at (-4.25, 2.75) {};
		\node [style=none] (13) at (-4.25, -2.75) {};
		\node [style=tightlabelnode] (14) at (-5, -2.75) {\small{$M_N$}};
		\node [style=tightlabelnode] (15) at (-5, 2.75) {\small{$M_1$}};
		\node [style=stateLarge] (16) at (-2.75, 0) {$\rho$};
		\node [style=tightlabelnode] (17) at (-5, 0) {$\goodvdots$};
		\node [style=none, doubled] (18) at (-0.5, -2.75) {};
		\node [style=none, doubled] (19) at (-0.5, 2.75) {};
	\end{pgfonlayer}
	\begin{pgfonlayer}{edgelayer}
		\draw [style=dashed] (2.center) to (5.center);
		\draw [style=dashed] (4.center) to (3.center);
		\draw [style=dashed, line width=5 pt, draw=white, in=180, out=0] (1.center) to (7);
		\draw [style=dashed, line width=5 pt, draw=white, in=180, out=0] (0.center) to (6);
		\draw [style=-, in=180, out=0] (1.center) to (7);
		\draw [style=-, in=180, out=0] (0.center) to (6);
		\draw [style=dashed] (12.center) to (19.center);
		\draw [style=dashed] (13.center) to (18.center);
	\end{pgfonlayer}
\end{tikzpicture}
\end{equation} 
In the context of a Bell-type measurement scenario, the processes $B_1,...,B_N$ are often referred to as \textbf{measurements}, their inputs as \textbf{measurement choices} and their outputs as \textbf{(measurement) outcomes}. We also refer to $\rho$ as the \textbf{shared state}, to $\mathcal{M} := \prod_{j=1}^N M_j$ as the set of \textbf{joint measurement choices} and to $\prod_{j=1}^N O_j$ as the set of \textbf{joint (measurement) outcomes}.

Note that our definition is purely a matter of \inlineQuote{shape}, and makes no direct reference to probabilistic or no-signalling structure: one of the points of strength of our framework is indeed in providing that structure automatically, based on the shape of the process alone. Indeed, the Bell-type measurement scenarios defined above provide a direct point of contact with the sheaf-theoretic framework for non-locality and contextuality \cite{Abramsky2011,Abramsky2015} (in a way similar to, but much more direct than, what done in \cite{Coecke2016,Chiribella2016,Fritz2014}). In the sheaf-theoretic framework, a non-locality scenario is specified by a \textbf{no-signalling empirical model} $(\zeta_{\,\underline{m}})_{\underline{m} \in \mathcal{M}}$, which is a family of $R$-distributions $\zeta_{\,\underline{m}}$ on the set $\prod_{j=1}^N O_j$ of joint outcomes indexed by the joint measurement choices $\underline{m} \in \mathcal{M}$ (here $R$ is the commutative semiring chosen to model the desired notion of non-determinism). 

\begin{theorem}[\textbf{Bell-type measurement scenarios}]\label{thm_BellTests}\hfill\\
A Bell-type measurement scenario in an $R$-probabilistic theory always yields a no-signalling empirical model in the sheaf-theoretic framework, with $R$ as commutative semiring modelling non-determinism.
\end{theorem}

This result shows that the sheaf-theoretic framework can be applied directly to the treatment of non-locality in $R$-probabilistic theories, and conversely that $R$-probabilistic theories are a natural environment to study the realisability of no-signalling empirical models from the sheaf-theoretic framework.

\section{Conclusions and future work}
\label{section_conclusions}

\subsection{Conclusions}

In this work, we have presented a general framework for probabilistic theories based upon three fundamental operational requirements: the existence of classical systems, the possibility of forming probabilistic combinations of processes (linear structure), and the possibility of discarding sub-systems (environment structure). Our requirements have a simple compositional and categorical formulation, as close as possible to the spirit and practice of Categorical Quantum Mechanics (CQM), and yet feature many of the fundamental structures used in Operational Probabilistic Theories (OPTs).

We hope that our work will contribute to providing a much-needed common ground for the CQM and OPT communities to collaborate on. In one direction, it brings the simplicity of CQM methods to OPTs, making it easy to create novel probabilistic theories, and to study existing ones from a fresh perspective. In the other direction, it allows methods and results from the OPT community to be straightforwardly applied to problems in CQM. 

Furthermore, we have shown that there is a direct connection between Bell tests in $R$-probabilistic theories and no-signalling empirical models in the sheaf-theoretic framework for non-locality and contextuality. As a consequence, methods from the sheaf-theoretic framework can be directly applied to problems of interest in CQM and OPTs, and conversely categorical methods can be used to study the realisability of empirical models in the sheaf-theoretic community.  

\subsection{Future work}

Firstly, we have provided a new point of view on the emergence of classical systems within a probabilistic theory $\CategoryC$ (and within quantum theory in particular) based upon the \textit{normalised} Karoubi envelope $\normalisedKaroubiEnvelope{\CategoryC}$. However, only a small subset of the objects of $\normalisedKaroubiEnvelope{\CategoryC}$ features in the construction. For example, it is known that all super-selected quantum systems appear in quantum theory by splitting all self-adjoint idempotent processes (i.e. using decoherence maps for both commutative and non-commutative special $\dagger$-Frobenius algebras) \cite{Selinger2008}, and this construction can be straightforwardly turned into a probabilistic theory. However, little or nothing is known about the operational significance of splitting general idempotent processes, and a thorough investigation will be the topic of future work.

Secondly, a number of theories independently developed in CQM and OPTs appear to be closely related, and should be properly connected. As a concrete example, we would be interest in applying our work to understand the connection between the ZW calculus \cite{Hadzihasanovic2015} on the CQM side and Fermionic quantum theory \cite{Fermionic1,Fermionic2} on the OPT side. Similarly, work on thermodynamics has been undertaken by both communities \cite{Bar2014,Chiribella2015,Chiribella2015a,TowardsThermo,Purity}, and should be appropriately related. 

Thirdly, we foresee a future, extended version of this work to contain a complete side-by-side translation of all common axioms from OPTs and constructions from CQM, for ease of use. The causality axiom has already been discussed, and two extremely important additional cases, namely the purity axioms from OPTs and the inner product structure from CQM, are covered in the Appendix.

Finally, some applications of our framework to the sheaf-theoretic study of non-locality and to quantum cryptography have recently appeared \cite{Gogioso2016e}. In the future, we expect similar constructions to provide new insight into other families of empirical models of interest to the sheaf-theoretic and quantum cryptography communities. Similarly, the framework has recently found application in the modelling of numerous toy theories of interest in quantum foundations \cite{Gogioso2017d}, and we expect this zoo to grow steadily in the coming years.

\newpage
\subparagraph*{Acknowledgements.}
The authors would like to thank Sean Tull, Matty Hoban and John Selby for comments, suggestions and useful discussions, as well as Sukrita Chatterji and Nicol\`o Chiappori for their support. SG gratefully acknowledges funding from EPSRC and the Williams Scholarship offered by Trinity College. CMS gratefully acknowledges funding from EPSRC and the Oxford-Google Deepmind Graduate Scholarship.


\newpage
\appendix

\section{Inner product structure} 

The category $\RMatCategory{R}$ always comes with compact closed structure, which takes the exact same form of the complex one from $\fdHilbCategory \simeq \RMatCategory{\complexs}$. It can similarly be endowed with dagger structure, as long as the semiring $R$ is involutive, i.e. it comes equipped with a self-inverse semiring homomorphism $\dagger: R \rightarrow R$. Every semiring can be turned into an involutive semiring by taking $\dagger := \id{R}$, and when talking about $\reals^+$ in the context of a dagger structure we will give as understood that the involution is the identity.

Pure-state quantum theory notably comes with a non-degenerate inner product structure, which features heavily in the traditional formalisms. In the context of CQM, the inner product structure of pure-state quantum theory is captured by the notion of $\dagger$-SMC, and intervenes in the construction of categories of completely positive maps \cite{Selinger2007,Coecke2012d}, the abstract counterparts to the operator model for mixed-state quantum theory. The non-degeneracy condition, on the other hand, can be captured directly in terms of the trace on positive operators, with no direct reference to the inner product structure itself: it says that the only positive operator $\rho$ satisfying $\Trace{\rho} = 0$ is the zero operator, and more generally that the only CP map $f$ satisfying $\Trace{f} = 0$ is the zero CP map. Because the trace provides the discarding maps for quantum theory, we will say that an $R$-probabilistic theory is \textbf{non-degenerate} if the following condition holds:
\begin{equation}\label{nonDegeneracy}
\begin{tikzpicture}[scale=0.8]
	\begin{pgfonlayer}{nodelayer}
		\node [style=box] (0) at (-9.5, 0) {$f$};
		\node [style=labelnode, inner sep=0.1 mm] (1) at (-12, 0) {};
		\node [style=trace] (2) at (-7, 0) {};
		\node [style=labelnode] (3) at (-5.5, 0) {$=$};
		\node [style=labelnode, inner sep=0.1 mm] (4) at (-4, 0) {$0$};
		\node [style=labelnode] (5) at (0, 0) {if and only if};
		\node [style=labelnode] (6) at (10.5, 0) {$=$};
		\node [style=labelnode, inner sep=0.1 mm] (7) at (12, 0) {$0$};
		\node [style=box] (8) at (6.5, 0) {$f$};
		\node [style=labelnode, inner sep=0.1 mm] (9) at (4, 0) {};
		\node [style=tightlabelnode] (10) at (9, 0) {};
	\end{pgfonlayer}
	\begin{pgfonlayer}{edgelayer}
		\draw [-] (1) to (0);
		\draw [style=-] (0) to (2);
		\draw [-] (9) to (8);
		\draw [style=-] (8) to (10);
	\end{pgfonlayer}
\end{tikzpicture}
\end{equation} 
A necessary condition for non-degeneracy of an $R$-probabilistic theory is that the classical theory be itself non-degenerate, and the latter requirement is equivalent to positivity of the semiring $R$ of scalars.
\begin{lemma}\label{lem_nonDegeneracy}
The classical theory $\RMatCategory{R}$ is non-degenerate if and only if $R$ is a positive semiring.
\end{lemma}

\section{Purity}
\label{section_purity} 

By a pure process we mean one which cannot involve any non-trivial interaction with a discarded environment. To be precise, we will say that a process $g$ is \textbf{pure} if the following implication always holds, for some normalised state $\rho$ (which will, in general, depend on $f$) \cite{Chiribella-pure}:
\begin{equation}\label{purity}
\begin{tikzpicture}
	\begin{pgfonlayer}{nodelayer}
		\node [style=box] (0) at (-6, 0) {$f$};
		\node [style=trace] (1) at (-3.75, -1.25) {};
		\node [style=none] (2) at (-5.5, -0.25) {};
		\node [style=tightlabelnode] (3) at (-3, 0.25) {};
		\node [style=tightlabelnode] (4) at (-8, 0) {};
		\node [style=none] (5) at (-5.5, 0.25) {};
		\node [style=none] (6) at (-6.5, 0) {};
		\node [style=tightlabelnode] (7) at (-9.5, 0) {$=$};
		\node [style=none] (8) at (-12.5, 0) {};
		\node [style=none] (9) at (-13.5, 0) {};
		\node [style=tightlabelnode] (10) at (-11, 0) {};
		\node [style=box] (11) at (-13, 0) {$g$};
		\node [style=tightlabelnode] (12) at (-15, 0) {};
		\node [style=tightlabelnode] (13) at (-0.5, 0) {$\Rightarrow$};
		\node [style=box] (14) at (11.5, 0) {$g$};
		\node [style=tightlabelnode] (15) at (9.5, 0) {};
		\node [style=tightlabelnode] (17) at (6.5, -1.25) {};
		\node [style=none] (18) at (4.5, 0.25) {};
		\node [style=tightlabelnode] (19) at (8, 0) {$=$};
		\node [style=none] (20) at (12, 0) {};
		\node [style=box] (21) at (4, 0) {$f$};
		\node [style=tightlabelnode] (22) at (13.5, 0) {};
		\node [style=none] (23) at (3.5, 0) {};
		\node [style=tightlabelnode] (24) at (6.5, 0.25) {};
		\node [style=tightlabelnode] (25) at (2, 0) {};
		\node [style=none] (26) at (4.5, -0.25) {};
		\node [style=none] (27) at (11, 0) {};
		\node [style=state] (28) at (11.5, -1.5) {$\rho$};
		\node [style=tightlabelnode] (29) at (13.5, -1.5) {};
	\end{pgfonlayer}
	\begin{pgfonlayer}{edgelayer}
		\draw [style=-] (3) to (5.center);
		\draw [style=-] (4) to (6.center);
		\draw [style=-, in=180, out=0] (2.center) to (1);
		\draw [style=-] (10) to (8.center);
		\draw [style=-] (12) to (9.center);
		\draw [style=-] (24) to (18.center);
		\draw [style=-] (25) to (23.center);
		\draw [style=-, in=180, out=0] (26.center) to (17);
		\draw [style=-] (22) to (20.center);
		\draw [style=-] (15) to (27.center);
		\draw [style=-, in=180, out=0] (28) to (29);
	\end{pgfonlayer}
\end{tikzpicture}
\end{equation} 
We will say that a $R$-probabilistic theory satisfies the \textbf{purification axiom} if every process $g $ can be written in the following form for some pure process $f$ (we refer to $\mathcal{E}$ as the \textbf{environment}) \cite{Chiribella2014}:
\begin{equation}\label{weakPurification}
\begin{tikzpicture}
	\begin{pgfonlayer}{nodelayer}
		\node [style=tightlabelnode] (0) at (0, 0) {$=$};
		\node [style=none] (1) at (-3, 0) {};
		\node [style=none] (2) at (-4, 0) {};
		\node [style=tightlabelnode] (3) at (-1.5, 0) {};
		\node [style=box] (4) at (-3.5, 0) {$g$};
		\node [style=tightlabelnode] (5) at (-5.5, 0) {};
		\node [style=none] (6) at (4, 0.25) {};
		\node [style=trace] (7) at (6.75, -0.75) {};
		\node [style=tightlabelnode] (8) at (1.5, 0) {};
		\node [style=none] (9) at (4, -0.25) {};
		\node [style=tightlabelnode] (10) at (5.5, -0.75) {$\mathcal{E}$};
		\node [style=none] (11) at (3, 0) {};
		\node [style=tightlabelnode] (12) at (7, 0.25) {};
		\node [style=box] (13) at (3.5, 0) {$f$};
	\end{pgfonlayer}
	\begin{pgfonlayer}{edgelayer}
		\draw [style=-] (3) to (1.center);
		\draw [style=-] (5) to (2.center);
		\draw [style=-] (12) to (6.center);
		\draw [style=-] (8) to (11.center);
		\draw [style=-, in=180, out=0] (10) to (7);
		\draw [style=-, in=180, out=0] (9.center) to (10);
	\end{pgfonlayer}
\end{tikzpicture} 
\end{equation} 
We will say that a probabilistic theory satisfies the \textbf{essentially unique purification axiom} \cite{Chiribella2010,ChiribellaNamely} if any two pure processes\footnote{Note that both processes have the same discarded environment system $\mathcal{E}$.} $f,g: \SpaceH \rightarrow \SpaceG \otimes \mathcal{E}$ satisfy the following implication, for some normalised invertible process $u: \mathcal{E} \rightarrow \mathcal{E}$ (which will, in general, depend on both $f$ and $g$):
\begin{equation}\label{e)up}
\begin{tikzpicture}
	\begin{pgfonlayer}{nodelayer}
		\node [style=box] (0) at (-6, 0) {$g$};
		\node [style=trace] (1) at (-3.75, -0.75) {};
		\node [style=none] (2) at (-5.5, -0.25) {};
		\node [style=tightlabelnode] (3) at (-3, 0.25) {};
		\node [style=tightlabelnode] (4) at (-8, 0) {};
		\node [style=none] (5) at (-5.5, 0.25) {};
		\node [style=none] (6) at (-6.5, 0) {};
		\node [style=tightlabelnode] (7) at (-9.5, 0) {$=$};
		\node [style=tightlabelnode] (8) at (-0.5, 0) {$\Rightarrow$};
		\node [style=tightlabelnode] (10) at (6.5, -0.75) {};
		\node [style=none] (11) at (4.5, 0.25) {};
		\node [style=tightlabelnode] (12) at (8, 0) {$=$};
		\node [style=box] (13) at (4, 0) {$f$};
		\node [style=none] (14) at (3.5, 0) {};
		\node [style=tightlabelnode] (15) at (6.5, 0.25) {};
		\node [style=tightlabelnode] (16) at (2, 0) {};
		\node [style=none] (17) at (4.5, -0.25) {};
		\node [style=none] (18) at (-13.5, 0.25) {};
		\node [style=box] (19) at (-14, 0) {$f$};
		\node [style=tightlabelnode] (20) at (-11, 0.25) {};
		\node [style=none] (21) at (-13.5, -0.25) {};
		\node [style=trace] (22) at (-11.75, -0.75) {};
		\node [style=none] (23) at (-14.5, 0) {};
		\node [style=tightlabelnode] (24) at (-16, 0) {};
		\node [style=box] (25) at (11.5, 0) {$g$};
		\node [style=none] (26) at (11, 0) {};
		\node [style=none] (27) at (12, -0.25) {};
		\node [style=none] (28) at (12, 0.25) {};
		\node [style=tightlabelnode] (29) at (9.5, 0) {};
		\node [style=tightlabelnode] (30) at (16, 0.25) {};
		\node [style=box] (31) at (14, -0.75) {$u$};
		\node [style=tightlabelnode] (32) at (16, -0.75) {};
	\end{pgfonlayer}
	\begin{pgfonlayer}{edgelayer}
		\draw [style=-] (3) to (5.center);
		\draw [style=-] (4) to (6.center);
		\draw [style=-, in=180, out=0] (2.center) to (1);
		\draw [style=-] (15) to (11.center);
		\draw [style=-] (16) to (14.center);
		\draw [style=-, in=180, out=0] (17.center) to (10);
		\draw [style=-] (20) to (18.center);
		\draw [style=-] (24) to (23.center);
		\draw [style=-, in=180, out=0] (21.center) to (22);
		\draw [style=-] (30) to (28.center);
		\draw [style=-] (29) to (26.center);
		\draw [style=-, in=180, out=0] (27.center) to (31);
		\draw [style=-] (31) to (32);
	\end{pgfonlayer}
\end{tikzpicture}
\end{equation} 

From a categorical perspective, it would be nice to phrase the \textit{purity preservation} axiom from OPTs as the statement that pure processes form a sub-SMC, but unfortunately we encounter a difficulty with the explicit presence of classical systems in the theory. Indeed, the identities on non-trivial classical systems are not pure, as they can be obtained by discarding a branch of a classical copy operation. As part of future work, we endeavour to consider more sophisticated notions of purity \cite{Tull2017,Selby2017,Selby2017leaks,Cunningham2017}, in order to bypass this issue and arrive at a suitable formulation of the purity preservation principle within our framework.

\section{Proofs} 

\subsection*{Proof of Lemma \ref{lem_SplitProbabilisticTheory} (p.\pageref{lem_SplitProbabilisticTheory})}

Because $\CategoryC$ is a full sub-SMC of $\normalisedKaroubiEnvelope{\CategoryC}$, we already have classical systems available, and all we need to show is that the enrichment and environment structure can be suitably extended. The extension of the enrichment is taken care of by the observation that any linear combination of processes invariant under idempotents is itself invariant. Indeed, the distributivity of linear structure over composition in $\CategoryC$ yields:
\begin{equation}
\Big[\forall j.\; g \circ f_j \circ h = f_j\Big] \Rightarrow \Big[g \circ (\sum_{j} p_j f_j) \circ h = \sum_{j} p_j f_j\Big]
\end{equation}  
Furthermore, the same discarding maps from $\CategoryC$ yield an environment structure for $\normalisedKaroubiEnvelope{\CategoryC}$: as the latter was defined in terms of \textit{normalised} idempotents, the discarding maps themselves are always invariant. 
\qed

\subsection*{Proof of Theorem \ref{thm_decoheredSystems} (p.\pageref{thm_decoheredSystems})}

All that we really need to show is that the process $f: (\SpaceH, \decoh{p,m}) \rightarrow (\SpaceG, \decoh{q,n})$ can be recovered from the classical process depicted in \ref{decoheredProcessesClassical}, and vice versa that any classical process $F: \classicalSystem{p,m} \rightarrow \classicalSystem{q,n}$ takes the form depicted in \ref{decoheredProcessesClassical} for a unique process $f: (\SpaceH, \decoh{p,m}) \rightarrow (\SpaceG, \decoh{q,n})$. The following picture proves both directions:
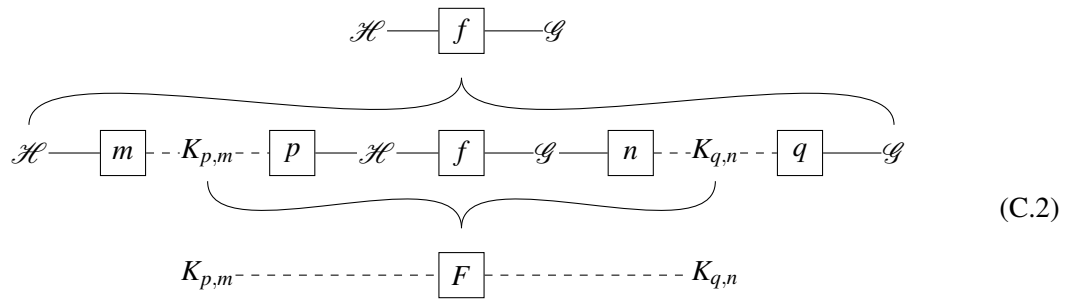
\begin{equation}\label{KaroubiDecohInvariantProcesses}
\begin{tikzpicture}
	\begin{pgfonlayer}{nodelayer}
		\node [style=box] (0) at (-10, 1.5) {$m$};
		\node [style=tightlabelnode] (1) at (-12.5, 1.5) {$\SpaceH$};
		\node [style=box] (2) at (-5.5, 1.5) {$p$};
		\node [style=tightlabelnode] (3) at (-3.25, 1.5) {$\SpaceH$};
		\node [style=box] (4) at (-1, 1.5) {$f$};
		\node [style=box] (5) at (3.5, 1.5) {$n$};
		\node [style=tightlabelnode] (6) at (1.25, 1.5) {$\SpaceG$};
		\node [style=box] (7) at (8, 1.5) {$q$};
		\node [style=tightlabelnode] (8) at (10.5, 1.5) {$\SpaceG$};
		\node [style=tightlabelnode] (9) at (-7.75, 1.5) {$\classicalSystem{p,m}$};
		\node [style=tightlabelnode] (10) at (5.75, 1.5) {$\classicalSystem{q,n}$};
		\node [style=none] (11) at (-7.75, 0.75) {};
		\node [style=none] (12) at (5.75, 0.75) {};
		\node [style=none] (13) at (-1, -0.5) {};
		\node [style=box] (14) at (-1, -1.75) {$F$};
		\node [style=tightlabelnode] (15) at (-7.75, -1.75) {$\classicalSystem{p,m}$};
		\node [style=tightlabelnode] (16) at (5.75, -1.75) {$\classicalSystem{q,n}$};
		\node [style=none] (17) at (-12.5, 2.25) {};
		\node [style=none] (18) at (10.5, 2.25) {};
		\node [style=none] (19) at (-1, 3.5) {};
		\node [style=box] (20) at (-1, 4.75) {$f$};
		\node [style=tightlabelnode] (21) at (1.5, 4.75) {$\SpaceG$};
		\node [style=tightlabelnode] (22) at (-3.5, 4.75) {$\SpaceH$};
	\end{pgfonlayer}
	\begin{pgfonlayer}{edgelayer}
		\draw [style=-] (1) to (0);
		\draw [style=-] (2) to (3);
		\draw [style=-] (3) to (4);
		\draw [style=-] (6) to (5);
		\draw [style=-] (7) to (8);
		\draw [style=-] (4) to (6);
		\draw [style=dashed] (0) to (9);
		\draw [style=dashed] (9) to (2);
		\draw [style=dashed] (5) to (10);
		\draw [style=dashed] (10) to (7);
		\draw [style=-, in=90, out=-90, looseness=0.75] (11.center) to (13.center);
		\draw [style=-, in=-90, out=90, looseness=0.75] (13.center) to (12.center);
		\draw [style=dashed] (15) to (14);
		\draw [style=dashed] (14) to (16);
		\draw [style=-, in=-90, out=90, looseness=0.50] (17.center) to (19.center);
		\draw [style=-, in=90, out=-90, looseness=0.50] (19.center) to (18.center);
		\draw [style=-] (22) to (20);
		\draw [style=-] (20) to (21);
	\end{pgfonlayer}
\end{tikzpicture}
\end{equation} 
\qed

\subsection*{Proof of Theorem \ref{thm_BellTests} (p.\pageref{thm_BellTests})}

We need to show two things: 
\begin{enumerate}
	\item[(i)] that each measurement choice $\underline{m} \in \prod_{j=1}^N M_j$ corresponds to an $R$-distribution on the joint measurement outcome set $\prod_{j=1}^N O_j$ (i.e. that the Bell-type measurement scenario corresponds to a well-defined conditional $R$-distribution);
	\item[(ii)] that marginalising over the outcome set $O_j$ for the $j$-th party yields an empirical model for the remaining parties which is independent of the individual measurement choice $m_j$ made by the $j$-th party herself (i.e. that the conditional $R$-distribution satisfies the no-signalling condition). 
\end{enumerate}
We begin by showing that the no-signalling condition (ii) holds, and we obtain condition (i) as a corollary. Marginalising over the $j$-th outcome set in an $R$-probabilistic theory corresponds to hitting the corresponding classical system $O_j$ with a discarding map:
\begin{equation}\label{BellTestproof1}
\begin{tikzpicture}
	\begin{pgfonlayer}{nodelayer}
		\node [style=none] (0) at (-10.25, 1) {};
		\node [style=none] (1) at (-10.25, -1.75) {};
		\node [style=none] (2) at (-6.5, 3) {};
		\node [style=none] (3) at (-3, -3) {};
		\node [style=none] (4) at (-6.5, -3) {};
		\node [style=none] (5) at (-3, 3) {};
		\node [style=box] (6) at (-7, 3) {$B_1$};
		\node [style=box] (7) at (-7, -3) {$B_{N}$};
		\node [style=tightlabelnode] (8) at (-7, 1.5) {$\goodvdots$};
		\node [style=tightlabelnode] (9) at (-2.25, 3) {\small{${O_1}$}};
		\node [style=tightlabelnode] (10) at (-2.25, -3) {\small{${O_N}$}};
		\node [style=tightlabelnode] (11) at (-2.25, 1.5) {$\goodvdots$};
		\node [style=trace] (12) at (-5, 1.75) {};
		\node [style=none] (13) at (-6.5, 2.75) {};
		\node [style=none] (14) at (-6.5, -3.25) {};
		\node [style=trace] (15) at (-5, -4.25) {};
		\node [style=none] (16) at (-10.25, 0) {};
		\node [style=trace] (17) at (-9.25, 0) {};
		\node [style=state] (18) at (-13.5, 4.5) {$m_1$};
		\node [style=state] (19) at (-13.5, -2.5) {$m_N$};
		\node [style=none] (20) at (-7.5, -3.25) {};
		\node [style=none] (21) at (-7.5, 3.25) {};
		\node [style=stateLarge] (22) at (-11, 0) {$\rho$};
		\node [style=tightlabelnode] (23) at (-13.5, -0.5) {$\goodvdots$};
		\node [style=none] (24) at (-10.5, -3.25) {};
		\node [style=none] (25) at (-10.5, 3.75) {};
		\node [style=tightlabelnode] (26) at (0, 0) {$=$};
		\node [style=none] (27) at (-6.5, -0.25) {};
		\node [style=trace] (28) at (-5, -1.25) {};
		\node [style=none] (29) at (-7.5, 0.25) {};
		\node [style=box] (30) at (-7, 0) {$B_{j}$};
		\node [style=state] (31) at (-13.5, 1.5) {$m_j$};
		\node [style=none] (32) at (-10.5, 2.75) {};
		\node [style=effect] (33) at (-2.25, 0) {$o_j$};
		\node [style=none] (34) at (-6.5, 0) {};
		\node [style=tightlabelnode] (35) at (-7, -1.5) {$\goodvdots$};
		\node [style=tightlabelnode] (36) at (-2.25, -1.5) {$\goodvdots$};
		\node [style=tightlabelnode] (37) at (-13.5, 3) {$\goodvdots$};
		\node [style=tightlabelnode] (38) at (-16, 0) {$\sum\limits_{o_j \in O_j}$};
		\node [style=none] (39) at (-10.25, -1) {};
		\node [style=tightlabelnode] (40) at (13.75, -1.5) {$\goodvdots$};
		\node [style=none] (41) at (5.5, 3.75) {};
		\node [style=none] (42) at (13, -3) {};
		\node [style=none] (43) at (9.5, -3) {};
		\node [style=stateLarge] (44) at (5, 0) {$\rho$};
		\node [style=none] (45) at (9.5, -3.25) {};
		\node [style=none] (46) at (5.75, -1.75) {};
		\node [style=box] (47) at (9, 3) {$B_1$};
		\node [style=box] (48) at (9, -3) {$B_{N}$};
		\node [style=none] (49) at (8.5, 3.25) {};
		\node [style=state] (50) at (2.5, 4.5) {$m_1$};
		\node [style=none] (51) at (5.5, 2.75) {};
		\node [style=none] (52) at (9.5, -0.25) {};
		\node [style=box] (53) at (9, 0) {$B_{j}$};
		\node [style=tightlabelnode] (54) at (2.5, 3) {$\goodvdots$};
		\node [style=none] (55) at (9.5, 3) {};
		\node [style=none] (56) at (13, 3) {};
		\node [style=trace] (57) at (11, -4.25) {};
		\node [style=tightlabelnode] (58) at (2.5, -0.5) {$\goodvdots$};
		\node [style=none] (59) at (9.5, 2.75) {};
		\node [style=none] (60) at (8.5, 0.25) {};
		\node [style=none] (61) at (5.75, -1) {};
		\node [style=none] (62) at (5.75, 1) {};
		\node [style=trace] (63) at (6.75, 0) {};
		\node [style=tightlabelnode] (64) at (13.75, -3) {\small{${O_N}$}};
		\node [style=trace] (65) at (11, 1.75) {};
		\node [style=trace] (66) at (11, -1.25) {};
		\node [style=tightlabelnode] (67) at (9, -1.5) {$\goodvdots$};
		\node [style=tightlabelnode] (68) at (13.75, 3) {\small{${O_1}$}};
		\node [style=none] (69) at (9.5, 0) {};
		\node [style=tightlabelnode] (70) at (13.75, 1.5) {$\goodvdots$};
		\node [style=trace] (71) at (13.75, 0) {};
		\node [style=state] (72) at (2.5, -2.5) {$m_N$};
		\node [style=none] (73) at (8.5, -3.25) {};
		\node [style=state] (74) at (2.5, 1.5) {$m_j$};
		\node [style=tightlabelnode] (75) at (9, 1.5) {$\goodvdots$};
		\node [style=none] (76) at (5.75, 0) {};
		\node [style=none] (77) at (5.5, -3.25) {};
	\end{pgfonlayer}
	\begin{pgfonlayer}{edgelayer}
		\draw [style=dashed] (2.center) to (5.center);
		\draw [style=dashed] (4.center) to (3.center);
		\draw [style=-, in=180, out=0] (13.center) to (12);
		\draw [style=-, in=180, out=0] (14.center) to (15);
		\draw [style=-, in=180, out=0] (16.center) to (17);
		\draw [style=dashed, line width=6 pt, draw=white, in=180, out=0] (1.center) to (7);
		\draw [style=dashed, line width=6 pt, draw=white, in=180, out=0] (0.center) to (6);
		\draw [style=-, in=180, out=0] (1.center) to (7);
		\draw [style=-, in=180, out=0] (0.center) to (6);
		\draw [style=dashed, in=180, out=0] (18) to (25.center);
		\draw [style=dashed, in=180, out=0] (19) to (24.center);
		\draw [style=dashed, in=180, out=0] (25.center) to (21.center);
		\draw [style=dashed] (24.center) to (20.center);
		\draw [style=dashed] (34.center) to (33);
		\draw [style=-, in=180, out=0] (27.center) to (28);
		\draw [style=dashed, in=180, out=0] (31) to (32.center);
		\draw [style=-, line width=5 pt, draw=white, in=180, out=0] (32.center) to (29.center);
		\draw [style=dashed, in=180, out=0] (32.center) to (29.center);
		\draw [in=180, out=0] (39.center) to (30);
		\draw [style=dashed] (55.center) to (56.center);
		\draw [style=dashed] (43.center) to (42.center);
		\draw [style=-, in=180, out=0] (59.center) to (65);
		\draw [style=-, in=180, out=0] (45.center) to (57);
		\draw [style=-, in=180, out=0] (76.center) to (63);
		\draw [style=dashed, line width=6 pt, draw=white, in=180, out=0] (46.center) to (48);
		\draw [style=dashed, line width=6 pt, draw=white, in=180, out=0] (62.center) to (47);
		\draw [style=-, in=180, out=0] (46.center) to (48);
		\draw [style=-, in=180, out=0] (62.center) to (47);
		\draw [style=dashed, in=180, out=0] (50) to (41.center);
		\draw [style=dashed, in=180, out=0] (72) to (77.center);
		\draw [style=dashed, in=180, out=0] (41.center) to (49.center);
		\draw [style=dashed] (77.center) to (73.center);
		\draw [style=dashed] (69.center) to (71);
		\draw [style=-, in=180, out=0] (52.center) to (66);
		\draw [style=dashed, in=180, out=0] (74) to (51.center);
		\draw [style=-, line width=5 pt, draw=white, in=180, out=0] (51.center) to (60.center);
		\draw [style=dashed, in=180, out=0] (51.center) to (60.center);
		\draw [in=180, out=0] (61.center) to (53);
	\end{pgfonlayer}
\end{tikzpicture}
\end{equation} 
But the measurements in a Bell-type measurement scenario are required to be normalised, and hence this corresponds to an empirical model for the remaining parties which is independent of the choice of classical deterministic state $m_j \in M_j$ (which is always normalised, and hence discarded to the scalar 1): 
\begin{equation}\label{BellTestproof2}
\begin{tikzpicture}
	\begin{pgfonlayer}{nodelayer}
		\node [style=none] (0) at (-10.25, 1) {};
		\node [style=none] (1) at (-10.25, -1.75) {};
		\node [style=none] (2) at (-6.5, 3) {};
		\node [style=none] (3) at (-3, -3) {};
		\node [style=none] (4) at (-6.5, -3) {};
		\node [style=none] (5) at (-3, 3) {};
		\node [style=box] (6) at (-7, 3) {$B_1$};
		\node [style=box] (7) at (-7, -3) {$B_{N}$};
		\node [style=tightlabelnode] (8) at (-7, 1.5) {$\goodvdots$};
		\node [style=tightlabelnode] (9) at (-2.25, 3) {\small{${O_1}$}};
		\node [style=tightlabelnode] (10) at (-2.25, -3) {\small{${O_N}$}};
		\node [style=tightlabelnode] (11) at (-2.25, 1.5) {$\goodvdots$};
		\node [style=trace] (12) at (-5, 1.75) {};
		\node [style=none] (13) at (-6.5, 2.75) {};
		\node [style=none] (14) at (-6.5, -3.25) {};
		\node [style=trace] (15) at (-5, -4.25) {};
		\node [style=none] (16) at (-10.25, 0) {};
		\node [style=trace] (17) at (-9.25, 0) {};
		\node [style=state] (18) at (-13.5, 4.5) {$m_1$};
		\node [style=state] (19) at (-13.5, -2.5) {$m_N$};
		\node [style=none] (20) at (-7.5, -3.25) {};
		\node [style=none] (21) at (-7.5, 3.25) {};
		\node [style=stateLarge] (22) at (-11, 0) {$\rho$};
		\node [style=tightlabelnode] (23) at (-13.5, -0.5) {$\goodvdots$};
		\node [style=none] (24) at (-10.5, -3.25) {};
		\node [style=none] (25) at (-10.5, 3.75) {};
		\node [style=tightlabelnode] (26) at (0, 0) {$=$};
		\node [style=none] (27) at (-6.5, -0.25) {};
		\node [style=trace] (28) at (-5, -1.25) {};
		\node [style=none] (29) at (-7.5, 0.25) {};
		\node [style=box] (30) at (-7, 0) {$B_{j}$};
		\node [style=state] (31) at (-13.5, 1.5) {$m_j$};
		\node [style=none] (32) at (-10.5, 2.75) {};
		\node [style=trace] (33) at (-2.25, 0) {};
		\node [style=none] (34) at (-6.5, 0) {};
		\node [style=tightlabelnode] (35) at (-7, -1.5) {$\goodvdots$};
		\node [style=tightlabelnode] (36) at (-2.25, -1.5) {$\goodvdots$};
		\node [style=tightlabelnode] (37) at (-13.5, 3) {$\goodvdots$};
		\node [style=none] (38) at (6, 0) {};
		\node [style=state] (39) at (2.75, 4.5) {$m_1$};
		\node [style=none] (40) at (13.25, 3) {};
		\node [style=tightlabelnode] (41) at (2.75, 3) {$\goodvdots$};
		\node [style=tightlabelnode] (42) at (14, -1.5) {$\goodvdots$};
		\node [style=tightlabelnode] (43) at (14, 3) {\small{${O_1}$}};
		\node [style=tightlabelnode] (44) at (14, 1.5) {$\goodvdots$};
		\node [style=trace] (45) at (11.25, -4.25) {};
		\node [style=trace] (46) at (11.25, 1.75) {};
		\node [style=tightlabelnode] (47) at (9.25, -1.75) {$\goodvdots$};
		\node [style=stateLarge] (48) at (5.25, 0) {$\rho$};
		\node [style=trace] (49) at (7, 0) {};
		\node [style=state] (50) at (2.75, 1.5) {$m_j$};
		\node [style=none] (51) at (13.25, -3) {};
		\node [style=trace] (52) at (9, 0.75) {};
		\node [style=tightlabelnode] (53) at (9.25, 1.75) {$\goodvdots$};
		\node [style=none] (54) at (9.75, 3) {};
		\node [style=state] (55) at (2.75, -2.5) {$m_N$};
		\node [style=none] (56) at (5.75, 3.75) {};
		\node [style=none] (57) at (8.75, -3.25) {};
		\node [style=none] (58) at (9.75, 2.75) {};
		\node [style=none] (59) at (8.75, 3.25) {};
		\node [style=none] (60) at (6, 1) {};
		\node [style=none] (61) at (6, -1.75) {};
		\node [style=tightlabelnode] (62) at (14, -3) {\small{${O_N}$}};
		\node [style=tightlabelnode] (63) at (2.75, -0.5) {$\goodvdots$};
		\node [style=none] (64) at (9.75, -3.25) {};
		\node [style=none] (65) at (5.75, 2.75) {};
		\node [style=box] (66) at (9.25, 3) {$B_1$};
		\node [style=none] (67) at (9.75, -3) {};
		\node [style=box] (68) at (9.25, -3) {$B_{N}$};
		\node [style=none] (69) at (5.75, -3.25) {};
		\node [style=none] (70) at (-10.25, -1) {};
		\node [style=none] (71) at (6, -1) {};
		\node [style=trace] (72) at (9, -0.5) {};
	\end{pgfonlayer}
	\begin{pgfonlayer}{edgelayer}
		\draw [style=dashed] (2.center) to (5.center);
		\draw [style=dashed] (4.center) to (3.center);
		\draw [style=-, in=180, out=0] (13.center) to (12);
		\draw [style=-, in=180, out=0] (14.center) to (15);
		\draw [style=-, in=180, out=0] (16.center) to (17);
		\draw [style=dashed, line width=6 pt, draw=white, in=180, out=0] (1.center) to (7);
		\draw [style=dashed, line width=6 pt, draw=white, in=180, out=0] (0.center) to (6);
		\draw [style=-, in=180, out=0] (1.center) to (7);
		\draw [style=-, in=180, out=0] (0.center) to (6);
		\draw [style=dashed, in=180, out=0] (18) to (25.center);
		\draw [style=dashed, in=180, out=0] (19) to (24.center);
		\draw [style=dashed, in=180, out=0] (25.center) to (21.center);
		\draw [style=dashed] (24.center) to (20.center);
		\draw [style=dashed] (34.center) to (33);
		\draw [style=-, in=180, out=0] (27.center) to (28);
		\draw [style=dashed, in=180, out=0] (31) to (32.center);
		\draw [style=-, line width=5 pt, draw=white, in=180, out=0] (32.center) to (29.center);
		\draw [style=dashed, in=180, out=0] (32.center) to (29.center);
		\draw [style=dashed] (54.center) to (40.center);
		\draw [style=dashed] (67.center) to (51.center);
		\draw [style=-, in=180, out=0] (58.center) to (46);
		\draw [style=-, in=180, out=0] (64.center) to (45);
		\draw [style=-, in=180, out=0] (38.center) to (49);
		\draw [style=dashed, line width=6 pt, draw=white, in=180, out=0] (61.center) to (68);
		\draw [style=dashed, line width=6 pt, draw=white, in=180, out=0] (60.center) to (66);
		\draw [style=-, in=180, out=0] (61.center) to (68);
		\draw [style=-, in=180, out=0] (60.center) to (66);
		\draw [style=dashed, in=180, out=0] (39) to (56.center);
		\draw [style=dashed, in=180, out=0] (55) to (69.center);
		\draw [style=dashed, in=180, out=0] (56.center) to (59.center);
		\draw [style=dashed] (69.center) to (57.center);
		\draw [style=dashed, in=180, out=0] (50) to (65.center);
		\draw [style=-, line width=5 pt, draw=white, in=180, out=0] (65.center) to (52);
		\draw [style=dashed, in=180, out=0] (65.center) to (52);
		\draw [in=180, out=0] (70.center) to (30);
		\draw [style=-, in=180, out=0] (71.center) to (72);
	\end{pgfonlayer}
\end{tikzpicture}
\end{equation} 
By repeating this process for all parties we are left with the scalar $\Trace{\rho}$: because the shared state $\rho$ in a Bell-type measurement scenario is required to be normalised, this is the scalar $1$. Hence the state over the joint measurement outcome set corresponding to the given joint measurement choice $\underline{m}$ was indeed normalised, i.e. we got an $R$-distribution as required by condition (i). 
\qed

\subsection*{Proof of Lemma \ref{lem_nonDegeneracy} (p.\pageref{lem_nonDegeneracy})}

Because we can always evaluate classical processes on deterministic input states, it is sufficient to show that the following two conditions are equivalent: (i) the zero state is the only classical state which yields the zero scalar when hit with the discarding map; (ii) the semiring $R$ of scalars is positive. A classical state on a classical system $X$ takes the form $p := \sum_{x \in X} p_x \delta_x$, where $\delta_x$ is the deterministic classical state concentrated at $x \in X$, and $(p_x)_{x \in X}$ is some family of scalars: as a consequence, $\Trace{p} = \sum_{x \in X} p_x$ is a generic sum of scalars. The requirement of positivity for the semiring $R$ states exactly that whenever $\sum_{x \in X} p_x = 0$ for some set $X$ and some family $(p_x)_{x \in X}$ of scalars, then $p_x = 0$ for all $x \in X$: as a consequence, the requirement of positivity is equivalent to asking that the equation $\Trace{p} = 0$ holds exactly when the classical state $p$ is the zero state $p := \sum_{x \in X} 0 \delta_x$.
\qed

\section{A brief compendium of fantastic quantum theories}

As mentioned in the introduction, the framework of categorical $R$-probabilistic theories can be used to capture a number of toy models of quantum theory: theories of wavefunctions valued in certain semirings, with classical systems emerging via a Born-like rule (embodied by the CP* construction). For the benefit of the reader, we present here a brief summary of the work done in \cite{Gogioso2017d}. 

If $S$ is any commutative semiring with involution, then we can defined the $\RMatCategory{S}$ as we did before, but now with additional dagger and compact closed structure given by the involution $^\ast$ on $S$. We can define the \textbf{sub-semiring of positive elements} $R$ for the semiring $S$ to be the closure under addition in $S$ of the set $\suchthat{x^\ast x}{ x \in S}$. Then it is possible to prove that $\CPStarCategory{\RMatCategory{S}}$ is an $R$-probabilistic theory under the $\CMonCategory$-enrichment inherited from $\RMatCategory{S}$. This result provides a template for generating quantum-like theories of wavefunctions valued in commutative involutive semirings, where classicality emerges via the same CP* construction used in traditional quantum theory to model the Born rule. A number of interesting quantum-like theories can be captured using this template:
\begin{enumerate}
	\item if $S$ is the complex numbers with complex conjugation, then $R = \reals^+$ is the probabilistic semiring and $\CPStarCategory{\RMatCategory{S}}$ is the probabilistic theory known as \textbf{quantum theory};
	\item if $S$ is the real numbers with the identity as involution, then $R = \reals^+$ is the probabilistic semiring and $\CPStarCategory{\RMatCategory{S}}$ is the probabilistic theory known as \textbf{real quantum theory};
	\item if $S$ is the split-complex numbers with split-complex conjugation as involution, then $R = \reals$ is the signed-probabilistic semiring and $\CPStarCategory{\RMatCategory{S}}$ is the quasi-probabilistic theory known as \textbf{hyperbolic quantum theory};
	\item if $S$ is the boolean semiring with the identity as involution, then $R$ is the boolean semiring and $\CPStarCategory{\RMatCategory{S}}$ is the possibilistic theory known as \textbf{relational quantum theory};
	\item if $S$ is a quadratic extension of the $p$-adic numbers with the conjugation from quadratic field extensions, then $R$ is the $p$-adic numbers and $\CPStarCategory{\RMatCategory{S}}$ is known as \textbf{$p$-adic quantum theory};
	\item if $S$ is a quadratic extension of a finite field $K$ with the conjugation from quadratic field extensions, then $R=K$ and $\CPStarCategory{\RMatCategory{S}}$ is a variation of \textbf{modal quantum theory}.
\end{enumerate}
In the future, we expect that the framework of $R$-probabilistic theories introduced in this work will be used to study these and other toy theories from a fully categorical perspective.

\end{document}